\documentclass[12pt,preprint]{emulateapj}
\usepackage{natbib,epsfig,wrapfig,floatflt,ulem}
\def\h2o{H$_2$O}
\def\ch4{CH$_4$}
\def\arcs{\ifmmode {''}\else $''$\fi}

\slugcomment{Draft copy \today}
\shorttitle{Physical Properties of Young Brown Dwarfs}
\shortauthors{Rice et al.}

\begin{document}

\title{Physical Properties of Young Brown Dwarfs and Very Low-Mass Stars Inferred from High-Resolution Model Spectra}

\author{{EMILY L. RICE\altaffilmark{1,2}, T. BARMAN\altaffilmark{3},
IAN S. MCLEAN\altaffilmark{1}, L. PRATO\altaffilmark{3}, J. DAVY KIRKPATRICK\altaffilmark{4}}}

\altaffiltext{1}{Department of Physics and Astronomy, UCLA, Los
Angeles, CA 90095-1562}
\altaffiltext{2}{Current address: American Museum of Natural History, Central Park West at 79th Street, New York, NY 10024-5192, erice@amnh.org}
\altaffiltext{3}{Lowell Observatory, 1400 West Mars Hill Road, Flagstaff, AZ 86001}
\altaffiltext{4}{Infrared Processing and Analysis Center, California Institute of Technology, Pasadena, CA 
91125}

\begin{abstract}
By comparing near-infrared spectra with atmospheric models, we infer the effective temperature, surface gravity, projected rotational velocity, and radial velocity for 21 very-low-mass stars and brown dwarfs. The unique sample consists of two sequences in spectral type from M6--M9, one of 5--10~Myr objects and one of $>$1~Gyr field objects. A third sequence is comprised of only $\sim$M6 objects with ages ranging from $<$1~Myr to $>$1~Gyr. Spectra were obtained in the $J$ band at medium (R$\sim$2,000) and high (R$\sim$20,000) resolutions with NIRSPEC on the Keck~II telescope. Synthetic spectra were generated from atmospheric structures calculated with the {\tt PHOENIX} model atmosphere code. Using multi-dimensional least-squares fitting and Monte Carlo routines we determine the best-fit model parameters for each observed spectrum and note which spectral regions provide consistent results. We identify successes in the reproduction of observed features by atmospheric models, including pressure-broadened K~{\sc i} lines, and investigate deficiencies in the models, particularly missing FeH opacity, that will need to be addressed in order to extend our analysis to cooler objects. The precision that can be obtained for each parameter using medium- and high- resolution near-infrared spectra is estimated and the implications for future studies of very low mass stars and brown dwarfs are discussed.
\end{abstract}


\section{\sc Introduction}
Determining the physical properties of brown dwarfs and very low-mass stars is important for our understanding of a broad range of topics including star and planet formation, circumstellar disks, dust formation in cool atmospheres, and the initial mass function. 
The direct measurement of mass and/or radius for a brown dwarf or very low-mass stars is possible only for certain binary systems, which are rare{\footnote{An archive of very-low-mass binaries is maintained by N.~Siegler (\url{http://vlmbinaries.org/); only a small fraction of these are double-lined spectroscopic binaries or visual binaries suitable for astrometric monitoring.}}. Determining other physical properties like effective temperature and surface gravity from bolometric luminosity estimates requires a precise distance measurement and assumptions about age (to determine effective temperature) plus mass and radius (to infer surface gravity from evolutionary models). Although radius is not expected to change very much for objects older than a few 100~Myr, reliance on evolutionary models is problematic as they are poorly calibrated at young ages. Commonly used conversions from spectral type to effective temperature (e.g. \citealt{Golimowski04} for field objects and \citealt{Luhman03} for young M dwarfs) rely on model isochrones and monotonic relationship between spectral type and effective temperature, which might not accurately represent the complicated formation and evolution of very-low-mass objects \citep[e.g.][]{Stassun06}. Evolutionary models also require atmosphere models to provide boundary conditions. Thus, synthetic spectra from such atmosphere models potentially offer a more direct method of inferring physical properties by comparison with observed spectra. 

The NIRSPEC Brown Dwarf Spectroscopic Survey (BDSS, \citealt{McLean03,McLean07}) is a large sample of high-quality near-infrared spectra obtained with the Keck~II 10-meter telescope. One of the goals of the BDSS is to provide as direct a measurement of effective temperature, surface gravity, and metallicity as is possible for non-eclipsing objects. In principle this can be accomplished by comparing observed infrared spectra with synthetic spectra from atmosphere models, without relying on spectral type, distance measurements, or estimates of radius, mass, and age. Other authors have applied this promising but as yet imperfect technique to brown dwarfs. The primary difficulty of applying this method to low-mass stars and brown dwarfs is the complexity of the spectra to be modeled, leading to imposed limitations on resolution, wavelength range, and/or sample size. For example, \citet{Mohanty04a,Mohanty04b} used narrow wavelength ranges at high resolution and \citet{Cushing08} used a broad wavelength range at low resolution. \citet{Reiners07} combined high-resolution spectra and broad wavelength coverage for a small sample of objects.

Many uncertainties remain in the determination of physical properties from the comparison of observed and synthetic spectra so it is important to establish the utility of this method using a larger sample of objects and a broader wavelength range. We accomplish this by drawing on medium- (R$\sim$2,000) and high-resolution (R$\sim$20,000) $J$-band spectra from the extensive BDSS data base{\footnote{A public archive is maintained by I.~S.~McLean and collaborators (\url{http://www.astro.ucla.edu/~mclean/BDSSarchive/)}}. Our goal is to show the extent to which synthetic spectra can be used to derive physical properties (effective temperature [T$_{eff}$], surface gravity [log(g)], projected rotational velocity [$v$sin$i$], and radial velocity [RV)]) rather than the measurement and analysis of trends in the observational data. This paper focuses on field ($\ge$~1~Gyr) and young (1 to $<$100~Myr) M dwarfs, the majority of which are predicted to be substellar. Future papers will deal with a larger sample of young M dwarfs, young and field L dwarfs, and finally T dwarfs. 

Section~2 describes the target selection, NIRSPEC observations, and data reduction. Section~3 outlines the calculation of atmospheric structures and synthetic spectra with the {\tt PHOENIX} model atmosphere code and our novel spectral fitting method. Section~4 presents the analysis of the spectral fits, including an overview of the results for each resolution and wavelength range and for the inferred physical properties. Our results are discussed in \S~5. Section~6 summarizes our conclusions and the implications of this work for future studies of brown dwarfs.

\section{\sc Observations and Data Reduction}
\subsection{Sample Selection}
The sample of 21 objects constitutes three overlapping subsamples: 1) young ($\sim$5--10~Myr) objects with spectral types from M6 to M8.5, 2) field objects ($\ge$~1~Gyr) with spectral types from M5.5 to M9, and 3) $\sim$M6 objects with ages from $\lesssim$1~Myr to $\ge$~1~Gyr. Targets were selected based on published estimates of spectral type (M5.5 to M9), brightness ($J<$~14.5), and age. The names, spectral types, coordinates, and $J$-band magnitudes of the sample are listed in Table~\ref{tbl-olog}. Often the sample contains more than one object of a given spectral type and age to look for possible differences between objects (effects of multiplicity, rapid rotation, etc.) and to aid in identifying fiducial objects with well-determined physical properties. Ages of young objects are determined from membership in near-by star-forming regions and moving groups or confirmed companionship to known young stars {age~$\sim$1~Myr to $\lesssim$100~Myr}, with the exception of 2MASS~0608$-$27, which has not been definitively associated with any other young objects. 
The young spectral type sequence contains objects with ages $\sim$5--10~Myr in order to avoid the complications introduced by extinction, ongoing accretion, and veiling. The age sequence of $\sim$M6 objects necessarily contains younger objects that show signs of accretion activity and higher extinction. The majority of our targets have $A_V$$\sim$0 with no evidence for ongoing accretion, although a few objects may retain cool disks. 
The majority of objects making up the 5--10 Myr sequence are confirmed low-mass members of the Upper Scorpius OB association, with additional young brown dwarfs from the $\sim$10~Myr-old TW Hydrae Association \citep[e.g.][]{Zuckerman04,BarNav06}. The sample properties are summarized in Table~\ref{samp}.

\subsection{Observations}
Observations were made between 2000 December and 2009 April as part of the NIRSPEC Brown Dwarf Spectroscopic Survey (BDSS, \citealt{McLean03,McLean07}). NIRSPEC is the cryogenic cross-dispersed echelle spectrometer on the Keck~II 10-m telescope on Mauna Kea, Hawaii \citep{McLean98,McLean00}. NIRSPEC has two modes, a cross-dispersed echelle mode and a non-echelle mode in which the echelle grating is replaced with a mirror and the cross-disperser provides the spectral dispersion. Both of these modes were employed with the NIRSPEC-3 (N3) filter, which approximates standard $J$-band coverage (1.143--1.375~$\mu$m). In echelle mode eight usable dispersion orders (65 to 58) are captured on the detector. Because the spectral interval captured by the detector is slightly smaller than the free spectral range in each order, there are small gaps, increasing with wavelength, in the total spectral coverage. The exact wavelength ranges for each dispersion order are listed in the headings in Setion~\ref{orders}. The slit width was two pixels (0\farcs38) for non-echelle observations and three pixels (0\farcs432) for echelle observations. The resolving power in $J$-band is approximately R=$\lambda$/$\Delta\lambda$=2,000 (``medium'' resolution) in non-echelle mode and R=20,000 (``high'' resolution) in echelle mode. Throughout the paper medium-resolution $J$-band spectra will be refered to by the filter name ``N3'' and high-resolution spectra by the number of the dispersion order.

Observing methods are described in detail in \citealt{McLean03} (medium resolution) and \citealt{McLean07} (high resolution); the following is a brief summary and explanation of departures from those methods. In both modes observations were made in pairs, nodding along the slit between each observation so that traces were separated by 20\arcsec~on the 42\arcsec-long slit (medium resolution) and by 7\arcsec~on the 12\arcsec-long slit (high resolution). For some recent high-resolution observations the size of the nod was reduced to avoid problems from an intermittent quadrant in the slit-viewing camera. During these occasions (2006 May and later) the nod size was at least 2\arcsec~so that the dispersed traces would be well-separated on the slit. Integration time was 600~seconds per nod for all echelle mode observations except those on 2006 May 19 and 20, which were 300~seconds per nod. Medium-resolution observations were typically 300~seconds per nod. Total integration times per object are listed in Table~\ref{tbl-olog}. A0~V stars were observed at an airmass very close to that of the target object (typically $<$0.1 airmass difference) to calibrate for telluric absorption features. Arc lamp spectra were obtained at least once per night, and white-light spectra and corresponding dark frames were obtained for flat-fielding.

\subsection{Data Reduction Methods}
All of the observed data were reduced with the REDSPEC IDL-based software package\footnote{http://www2.keck.hawaii.edu/inst/nirspec/redspec.html}, described in \citet{McLean03,McLean07}. The package performs standard bad pixel interpolation, dark subtraction, and flat-fielding as well as spatial rectification of curved spectra. Spectra are rectified and extracted in differenced nod pairs so that the sky background and OH emission lines are removed. Spectra were extracted by summing over 7--15 pixels dependent on seeing then subtracted again to produce a positive spectrum with residual sky emission features removed. A0~V star spectra were reduced the same way, and target spectra were divided by the A0~V star spectrum to remove telluric features and multiplied with a T$_{eff}=9500$ K blackbody spectrum to restore the true slope. At medium resolution Pa$\beta$ absorption at $\lambda$=1.2822~$\mu$m in the A0~V star spectrum is removed by interpolation, and at high resolution order 59 was not divided by an A0~V as it lacks strong telluric lines. Medium-resolution spectra were wavelength calibrated using Ne and Ar arc lamp lines, and high-resolution spectra with OH night sky lines, which were found to be highly stable and well-distributed across orders. Seven high-resolution dispersion orders (58--65, excepting 60) were reduced. Order 60 was not reduced because the OH night sky lines are blended with O$_2$ emission bands at 1.26--1.28~$\mu$m \citep{Rousselot00} making wavelength calibration and sky subtraction considerably more difficult. Each reduced spectrum was continuum normalized, and multiple nod pairs were averaged together to increase SNR. High-resolution spectra were shifted to the heliocentric reference frame.

Because the sample objects were selected in order to minimize the effects of extinction, only GY~5 required correction for its relatively large extinction, $A_V$$\sim$5. The medium-resolution N3 spectrum was dereddened as described in \citet{McGovern05}. High-resolution observations were not dereddened because the spectra cover such narrow wavelength ranges and are considered individually in the analysis, making differences in the pseudo-continuum level caused by reddening relatively unimportant.

\section{\sc Model Atmospheres and Synthetic Spectra}
\subsection{{The \tt PHOENIX} Code} 
\label{phxcode}
The {\tt PHOENIX} code was developed by Peter Hauschildt and collaborators beginning about 20 years ago, originally to model radiative transfer in supernova remnants. More recent versions of the code have been successfully applied to modeling stellar, substellar, and planetary atmospheres \citep{Hauschildt97,Hauschildt99,Allard01,Barman01}. Some of the capabilities and features of {\tt PHOENIX} are reviewed by \citet{Baron03}. Atmospheres are constructed by calculating temperature and pressure in each of 64 spherically-symmetric layers. The {\tt PHOENIX} code determines radiative fluxes through the layers under the assumptions of hydrostatic equilibrium, radiative-convective equilibrium, and chemical equilibrium. The code begins with the results of a previously calculated structure, as an initial guess, and iterates toward the new solution by changing the radiative-convective temperatures at each layer via a modified Uns\"{o}ld-Lucy temperature correction procedure \citep{Hauschildt03}. Models were run for at least 20 iterations, which was sufficient in most cases to reach convergence. Convergence of a model is based on energy conservation and determined by an upper-limit on the percentage difference between the radiative-convective flux and the desired model flux ($\sigma$~T$_{eff}^4$), typically 5\% or less at all layers. The output for converged atmospheres includes spectral energy distributions (SEDs) from 10~\AA~ to 1~mm with a resolution of 4~\AA~through the optical and near-infrared. From a converged atmospheric structure the PHOENIX code is used to produce a synthetic spectrum for a selected wavelength range (i.e., the $J$ band) and numerical resolution, up to 500,000 wavelength elements in total. 

\subsection{Brown Dwarf Model Atmospheres}
\label{bdmodels}

To match our observational sample we calculate model atmospheres for effective temperatures from 1800~K to 3000~K in intervals of 50~K and surface gravities from log(g)~=~3.0 to 6.0 in intervals of 0.1~dex. For this temperature range the {\it dusty} version of the {\tt PHOENIX} model code is applicable. In this version, dust is formed in chemical equilibrium conditions and remains where it is formed in the atmosphere; the removal of dust by dynamical processes is ignored. At low temperatures where more dust is created (T$_{eff}$~$\le$~2200~K), the models often do not converge because of the drastic effects dust production has on the temperature structure. Instead, the temperature iterations will cause fluctuations around the converged solution. In these cases we manually chose the temperature-pressure structure with the smallest difference between calculated and prescribed flux ($\sigma$~T$_{eff}$), which was comparable to the differences for automatically converged models. All dust is created as pure, single-species, spherically-symmetric grains that follow an interstellar grain-size distribution \citep{Allard01}. Model calculations incorporate some updates to the \citet{Allard01} models, including solar abundances from \citet{Asplund05}, the FeH line list from \citet{Dulick03}, the CrH line list from \citet{Burrows02}, and the H$_2$O line list from \citet{Barber06}. The incorporation of the new FeH line list causes the most drastic change from previous generations of {\tt PHOENIX} models and is discussed in detail in \S~\ref{sFeH}. The differences are most evident in the medium- and high-resolution spectra and have little effect on the atmospheric structures. We integrate under the SEDs and compare the total flux to $\sigma$~T$_{eff}^4$, and the differences were typically less than 2\% but up to 5\% for models with T$_{eff}$~$<$~2000~K. Figure~\ref{models} shows the structures and SEDs for a range of effective temperatures at a surface gravity of log(g)=5.0 and for a range of surface gravities at T$_{eff}$=2600~K. The non-smoothness in the structures at cool temperatures and low pressures are induced by dust formation and are mainly above the near-infrared photosphere, which lies at approximately 10$^4$--10$^7$~dynes~cm$^{-2}$ depending on effective temperature and surface gravity.

Pressure-broadening of strong lines impacts both the structure and the emergent spectrum of an atmosphere by adding opacity at wavelengths up to thousands of angstroms outside of the line center \citep[e.g.][]{Burrows00,Allard01}. The numerical treatment of Van der Waals broadening in the {\tt PHOENIX} models is described in \citet{Schweitzer96}. A more detailed treatment of the line broadening incorporates multiple perturbers in calculating the line wings \citep{AllardN05}. We tested this treatment and found that the differences in structures, SEDs, and high-resolution spectra were minimal for the temperatures and surface gravities under consideration. We also experimented with how widely the K and Na line profiles are calculated from the line center. This is particularly important for alkali lines in the optical that have wings extending thousands of angstroms and contribute a large amount of opacity; when the line profiles are truncated it affects the atmospheric structure. The minimum width of the line profile above which there was not a noticeable difference was used. The consideration of line broadening will be more crucial for cooler atmosphere models (i.e. T dwarfs) for which dust opacity no longer dominates.

Medium-resolution synthetic spectra were calculated for the wavelength range 1.1--2.5~$\mu$m at 0.3~\AA~resolution to oversample the instrumental resolution for the non-echelle (medium resolution, R$\sim$2,000) mode of NIRSPEC at all near-infrared bands. For the current analysis only the $J$-band spectra are considered, but future papers will extend this study to $H$- and $K$-band BDSS data. High-resolution synthetic spectra were calculated at 0.03~\AA~resolution for the wavelength range 1.1--1.4~$\mu$m to cover the entire $J$~band and oversample the instrumental dispersion of NIRSPEC by a factor of 5--6 at those wavelengths \citep[][Table~2]{McLean07}. The output of the {\tt PHOENIX} code can be used to identify the most prevalent atomic, molecular, and dust grain opacity sources. 

\subsection{Spectral Fitting}
\label{fitting}
The MPFIT IDL code develeped by \citet{Markwardt09} was used to implement Levenberg-Marquardt least-squares minimization and determine the best-fit model parameters for each spectrum individually. The fit parameters are $T_{eff}$ and log(g) from the grid of atmosphere models. Projected rotational velocity ($v$sin$i$) is incorporated using standard Uns\"{o}ld-type profile and convolution. Radial velocity (RV) is implemented as a shift in the observed spectrum onto the model wavelength array. A subroutine of the fitting code linearly interpolates between calculated models for each flux value in a given wavelength range in order to draw from an essentially smooth model grid. 
Once a new model is created for $T_{eff}$ and log(g), the $v$sin$i$ kernel and RV correction are applied. The $v$sin$i$ kernel uses a limb-darkening coefficient of 0.6, which is typical although not well constrained. The flux values of synthetic spectra created using different limb-darkening coefficients differ by less than a tenth of one percent on average even at the lowest surface gravities and highest $v$sin$i$ values considered. 

The MPFIT routine was implemented two different ways. First, the initial $T_{eff}$ and log(g) parameters were randomly drawn from the continuous range covered by linear interpolation of the calculated synthetic spectra. Some of the resulting best-fit parameters had much higher $\chi^2$ values than was typical, indicating the code had found a local minimum rather than a global minimum, which is a known drawback of Levenberg-Marquardt least-squares minimization. The best-fit results were not sensitive to initial values of $v$sin$i$ and RV. The best-fit parameters with with lowest $\chi^2$ values were averaged and set as the initial parameters for the second implementation of MPFIT. Following the Monte Carlo method, the initial guesses were kept constant and each flux point of the observed spectrum was randomly resampled from within the noise on that pixel. The distribution of the best-fit parameters then indicates the uncertainty on the results from the noise in the data, which are adopted as our formal (relative) uncertainties, knowing that uncertainties in the model are much more difficult to characterize. Thus, like similar studies by \citet{Mohanty04a}, for example, our uncertainties represent relative precision and not accuracy. A veracious description of the uncertainties in the atmosphere models would require, for example, comparisons of results from different sets of models \citep[e.g.][]{Helling08b}, which is beyond the scope of this paper. Typical results of this procedure are illustrated in Figure~{\ref{chis}. All spectra were fit using 10$^4$ iterations of each implementation of MPFIT. Several spectra were fit with both 10$^3$ and 10$^4$ iterations, and the differences in resulting best-fit parameters were negligible.

The results for each spectral fit are presented in Tables~\ref{tbl-10myr-temp}--\ref{tbl-vel}. Tables~\ref{tbl-10myr-temp} and \ref{tbl-10myr-logg} present the best-fit parameters for effective temperature and surface gravity, respectively, of each N3 spectrum and high-resolution dispersion order for the spectral type sequences. Each result is coded according to the quality for all orders based on visual inspection. No annotation corresponds to a good fit with minimal mismatches between the observed and synthetic spectra. Poorer fits are denoted by ({\it value}) for a few mismatches in depth or wing-shape for the strongest lines and by -({\it value})- for several mismatches. Entries that are struck are very poor fits that were not used in determining the adopted best-fit values. Note that the annotations are the same for both tables, even though it might be the case that the gravity is acceptable and the temperature is causing the poor fit, or vice versa. The quality of fits for each order are described in \S~\ref{orders} and for each parameter in \S~\ref{objects}. Tables~\ref{tbl-M6age-temp} and \ref{tbl-M6age-logg} present the best-fit parameters for the $\sim$M6 objects and are annotated the same way. The {\it adopted} T$_{eff}$ and log(g) values are the mean of results from each dispersion order, weighted by the sum of the squared errors for all orders. With this method parameters that result in a good fit for one order but a poorer fit for other orders receive less weight. The velocity results presented in Table~\ref{tbl-vel} are determined by the same weighted-mean method. The results are discussed by parameter in \S~\ref{teff}--\ref{sVel}. The N3-fit results were not used in the determination of the adopted best-fit values.

\section{\sc Analysis}

In the following sections we discuss the medium-resolution (R$\sim$2,000, referred to as N3) and high-resolution (R$\sim$20,000, referred to by dispersion order) synthetic and observed spectra in terms of major temperature- and gravity-dependent features. Then we summarize the fitting results for each parameter ($T_{eff}$, log(g), $v$sin$i$, and RV). 

\subsection{Overview of Spectral Fits}
\label{orders}

{\it N3 (1.143 -- 1.375 $\mu$m)} --- The N3 filter on NIRSPEC corresponds to the $J$ band, the wavelength regime in which flux of M dwarfs peaks. We use ``N3'' as shorthand for medium-resolution $J$-band spectra throughout the following sections. The morphology of observed spectra for M~dwarfs with different spectral types and ages is shown in Figure \ref{medres3}. 
The overall shape of the N3 spectrum is sensitive to effective temperature, while the strengths of atomic lines are sensitive to surface gravity. At lower temperatures the pseudo-continuum shape becomes more dependent on surface gravity. For most objects the N3 fits are very good, with the exception of the region between 1.20 and 1.24 $\mu$m, which is likely missing FeH opacity in the atmosphere models (see \S~\ref{sFeH}), and at 1.28~$\mu$m where young and late-type objects might have weak Pa$\beta$ emission. For some objects the depth of the H$_2$O band starting at 1.335~$\mu$m is not well matched, with the band being too strong in the atmosphere models for later-type objects. The apparent mismatch in the H$_2$O depth could also be a results of a poor fit to the pseudo-continuum level just before the H$_2$O band. Qualitatively the medium-resolution synthetic spectra provide good fits to the observed spectra, but the best-fit parameters are often discrepant from values predicted by the evolutionary models of \citet{Chabrier00} and \citet{Baraffe02}.

{\it Order 65 (1.16496 -- 1.18207 $\mu$m)} --- The shortest-wavelength dispersion order features a strong K~{\sc i} doublet (resolved into a triplet at high resolution), a few weak Fe and Ti lines, and molecular lines of FeH and H$_2$O. The K~{\sc i} lines are sensitive to both temperature and surface gravity because both decreasing temperature and increasing surface gravity contribute to pressure-broadening (Figure~\ref{models65}). The strength of the molecular lines can break the degeneracy between high gravity and low temperature, particularly the strongest H$_2$O line at 1.16579~$\mu$m, which is weakly sensitive to gravity but strengthens greatly with decreasing temperature. The synthetic spectra reproduce the pressure-broadening of the K~{\sc i} lines for later spectral types (decreasing effective temperatures) and increasing age (increasing surface gravity) remarkably well. One difference between the order 65 observed and synthetic spectra is the shape of the pseudo-continuum between the K~{\sc i} lines at low temperatures. In the synthetic spectra the pseudo-continuum is more rounded as opposed to slightly flattened in the observed spectra. This is likely an indication that the alkali line profiles are not calculated as far into the wings as they should be. This was a known problem in the optical where the alkali lines are so strong that the calculation of the line profiles affects the atmospheric structure, but it is only a minor effect in the near-infrared.

The atmosphere models provide excellent fits to the order 65 observed spectra because of both the well-modeled K~{\sc i} lines and the lack of strong FeH lines in this particular wavelength region. Although even the strongest H$_2$O line is not strong enough to break the degeneracy between temperature and gravity with our fitting method, the linear relationship is so tight (see Figure~\ref{chis}, first panel) that constraining either temperature or gravity based on spectral type or evolutionary models, respectively, allows only a narrow range of values for the other parameter. In general the best-fit parameters from order 65 provide excellent fits to the other orders (with the notable exceptions of 62 and 63, see below), but in some cases the model fits produce K~{\sc i} lines that are too shallow in order 61 and Al {\sc i} lines that are too deep with wings that are too broad in order 59. This may however be a consequence of the underrepresented FeH opacity in the atmosphere models that is changing the relative level of the pseudo-continuum and not a failure of the best-fit model parameters.

{\it Order 64 (1.18293 -- 1.20011 $\mu$m)} --- Order 64 contains several atomic lines of Fe and Ti that are insensitive to temperature and gravity except at the low temperature/high gravity region of the parameter space covered in this analysis. This order provides less consistent fits for all but the earliest spectral type young objects because the FeH lines from the 0-1 band of F$^{4}$~$\Delta$-X$^{4}$~$\Delta$ at 1.1939~$\mu$m that strengthens at lower temperatures and higher gravities are not reproduced by the atmosphere models (see \S~\ref{sFeH}). Weaker atomic lines of Mg (1.1834 $\mu$m) and Ti (1.1896 \& 1.1953 $\mu$m) identified in the model spectra are present in the observed spectra, although they are more blended with other features in the observed spectra than the atmosphere models predict (Figure~\ref{models64}). Order 64 produces good fits for the earliest spectral type young objects, with the fits becoming worse with later spectral types, resulting in high temperatures and low surface gravities that produce too-weak K~{\sc i} lines in order 65 and 61 and too-strong Al~{\sc i} lines in order 58.

{\it Order 63 (1.20168 -- 1.21938 $\mu$m)} --- The only atomic lines in order 63 are very weak so that the spectral features are primarily FeH and H$_2$O. As mentioned above, the FeH lines in the atmosphere models are too weak for all but the earliest spectral types of young objects. Figure~\ref{models63} shows the general correspondence between absorption features and the increasing mismatch in their strengths at older ages and later spectral types. In both the models and the observations the strength of the FeH lines increases more dramatically with decreasing temperature and later spectral type than with decreasing gravity (the latter is more apparent at even lower temperatures than are shown in Figure~\ref{models63}). For young objects the correspondence between the locations of the lines, if not the strengths, is enough to provide a reasonable radial velocity measurement in most cases (see \S~\ref{sVel}).

{\it Order 62 (1.22093 -- 1.23899 $\mu$m)} --- Order 62 is also dominated by FeH lines that are not strong enough in the atmosphere models (Figure~\ref{models62}), particularly strong lines from the $Q$-branch of the $F^{4}$$\Delta_{7/2}$-$X^{4}$$\Delta_{7/2}$ system from 1.221 to 1.223 $\mu$m and $P$- and $R$-branch lines throughout the order \citep{McLean07}. Therefore order 62 is very similar to order 63 in terms of fitting observed spectra with synthetic spectra: for the most part the fit parameters are inconsistent except for radial velocity.

{\it Order 61 (1.24081 -- 1.25913 $\mu$m)} --- The longer-wavelength K~{\sc i} doublet in the $J$ band falls in order 61 (Figure~\ref{models61}). While the atmosphere models again replicate the dependence of the K~{\sc i} lines on temperature and gravity very well, the weaker lines in this order are more problematic than in order 65 for two reasons. First, the model predicts evenly-spaced TiO lines that increase in strength with decreasing temperature, particularly at low gravity, but such features are not seen in the observations. This is likely because TiO lines in the observed spectra are blended with FeH lines that are too weak in the atmosphere models. Second, the lack of strong FeH in the models causes the wings of the K~{\sc i} lines and the pseudo-continuum shape to not match the observations. Therefore order 61 provides less consistent model fits than order 65 despite the strong, well-modeled K~{\sc i} lines.

Furthermore, the best-fit parameters do not fit other orders as well as the fits from order 65 that are also anchored by K~{\sc i}. For all but the earliest spectral types (M6--M7), the best-fit parameters from order 61 result in Al {\sc i} lines that are too strong in order 58 for the young objects and for the field objects, an unrealistic combination of high temperature and high gravity or low temperature and low gravity, both of which are manifested as poor fits in the other orders.

{\it Order 59 (1.28262 -- 1.30151$\mu$m)} --- The atomic lines in order 59 are weak, but they have interesting dependence on temperature and gravity. The strongest lines are Ti and Mn, with Fe and Cr lines being too weak to distinguish from noise on the observational spectra. The strongest Ti lines (1.2835 and 1.2850 $\mu$m) and Mn line (1.2903 $\mu$m) become broader and shallower with decreasing temperature at high gravity in the atmosphere models, but the trend is not as apparent in the observed spectra of field objects (Figure~\ref{models59}). At low gravity the lines do not change much with temperature. At high gravity and low temperature order 59 also contains FeH lines that are not strong enough in the models, rendering this order poor for model fits. The narrow lines from 1.297 to 1.301 $\mu$m in the spectra of the young objects are weak telluric lines that were not removed because this order contains Pa$\beta$ in the spectrum of an A0~V star. Order 59 provides consistent fits only for the earliest spectral type young objects.

{\it Order 58 (1.30447 -- 1.32370$\mu$m)} --- The longest-wavelength order contains an Al doublet that, in the atmosphere models, is sensitive to temperature at high gravities but nearly unchanged with temperature at log(g)~$\le$~4.0 (Figure~\ref{models58}). The wings of the lines change shape with decreasing temperature as they become blended with strengthening molecular lines. This behavior is also seen in the high-gravity field objects, but not the young objects, for which the Al lines are stronger in the atmosphere models than in the observations. This is likely a consequence of the ubiquitous FeH absorption lowering the pseudo-continuum level making the Al lines appear relatively weaker than in the models, in which the FeH is too weak. Order 58 provides only slightly more consistent fits than order 59.

\subsection{Spectral Fitting Results}
\label{objects}

\subsubsection{Effective Temperature}
\label{teff}

Tables~\ref{tbl-10myr-temp} and \ref{tbl-M6age-temp} present previously determined effective temperatures from the literature (column 3), the effective temperatures derived from model fits to medium-resolution N3 spectra (column 4) and each of seven dispersion orders of high-resolution spectra (columns 5--11), and our adopted best-fit temperatures (column 12). Previously published effective temperatures are a combination of measurements of bolometric luminosity \citep[e.g.][]{Vrba04,Golimowski04}, spectral type to temperature conversions \citep[][]{Luhman03,Mohanty03b}, and comparison with synthetic spectra at medium resolution \citep[e.g.][]{McGovern05}. For many of the objects the range of temperatures reported in the literature is 300--400~K. Almost all of our adopted temperatures fall within or are very close to the range of previously determined temperatures; notable exceptions are TWA~5B and the field objects, discussed below. Adopted temperatures, which are from high-resolution fits, are also very similar to best-fit results from the N3 spectra, within $\sim$60~K for objects earlier than M7 with the exception of Gl~406, also discussed below. The results become less consistent for the later spectral type objects, with the spectral fits producing higher temperatures than predicted by spectral-type effective temperature relationships and bolometric measurements. For the most part the adopted temperatures decrease with later spectral types and are within $\sim$100~K for objects of the same spectral type, again with the notable exception of TWA~5B. Both the N3 and echelle spectra of TWA~5B are lower signal-to-noise than average for our observations and the spectra are likely contaminated by light from the $J$=7.67 mag primary $\sim$2.5\arcsec~away. Adopted effective temperatures for the field dwarf sequence also decrease for later spectral types, but they are even more dissimilar to previous temperature measurements.

The measured effective temperatures are most discrepant for the latest spectral type young objects, in particular TWA~5B, and the field (old, high gravity) objects Gl~406, LP~402-58, LP~412-31, and 2MASS~0140+27. For all of these objects the temperatures measured via spectral fitting are 200--350~K hotter than the effective temperatures determined by spectral type-effective temperature scales and bolometric luminosity measurements. The higher temperatures measured from spectral fitting are likely a consequence of the increasing importance of FeH absorption for cooler and higher gravity objects that is not reproduced by the atmosphere models (see \S~\ref{sFeH}). 

The age sequence of $\sim$M6 objects (Table~\ref{tbl-M6age-temp}) shows a very small spread in measured effective temperature. 
 Interestingly the hottest temperature was measured for the unresolved binary Gl~577BC, which is the earliest spectral type object in the subsample. The average temperature for the objects is 2880~K, with a standard deviation of 60~K, which, although cooler than the temperature of 2990~K for a young M6 on the \citet{Luhman03} scale, is promising for the development of a spectral type-effective temperature scale that is consistent for objects with a range of properties (e.g., age, rotation, and binarity) and calibrated by benchmark objects for which observed properties are successfully reproduced by atmosphere models. 

Uncertainty resulting from noise in the observed spectra is determined via the standard deviation of the best-fit effective temperatures from flux-re-sampled spectra, which is typically $\lesssim$~10~K but as high as 30~K. Systematic uncertainty in the accuracy of the atmosphere models likely dominates the total uncertainty in the results, although as discussed above uncertainties in the models are difficult to quantify (see \S~\ref{fitting}). The standard deviations of best-fit effective temperatures from 10$^4$ iterations with varied initial parameters are typically 100--200~K. This is similar to the standard deviation of the best-ft effective temperatures from all orders, which is additionally subject to varying levels of fidelity between the observations and the models (described in \S~\ref{orders}). The systematic uncertainties in effective temperature are much lower for the medium-resolution spectra, typically $\lesssim$~50~K. The implications of the relative uncertainties from high- and medium-resolution spectra are discussed in \S~\ref{diss1}.

\subsubsection{Surface Gravity}
\label{logg}

The surface gravity measurements are presented in Tables~\ref{tbl-10myr-logg}~and~\ref{tbl-M6age-logg}. Previous studies have discovered gravity-sensitive features and spectral indices in the near-infrared \citep[e.g.][]{Gorlova03,McGovern04,Allers07} with low (R$\sim$300) and medium (R$\sim$2,000) spectra, notably a sharply peaked $H$-band spectrum (from increased H$_2$O absorption shortward of the peak in $H$-band flux) and weaker atomic lines (K~{\sc i} and Na~{\sc i}). Our analysis differs from previous work in that surface gravity and effective temperature are determined simultaneously and at higher resolution. For young objects, the adopted log(g) from high-resolution fits is within 0.3~dex of the best-fit results for the medium-resolution spectra, with the exception of the TW Hydrae members and 2MASS~0608$-$27. For these objects the medium-resolution fits produced surface gravities nearly an order of magnitude lower than expected for their temperatures and ages based on the evolutionary models of \citet{Chabrier00} and \citet{Baraffe02} (hereafter DUSTY00), an indication that high-resolution spectra are better for examining gravity-sensitive features than medium-resolution spectra. For field objects the medium-resolution fits produce gravities that are higher than physically reasonable according to the DUSTY00 evolutionary models. Such high gravities are likely a result of the N3 fits being affected by the missing FeH opacity in the atmosphere models. The surface gravity measurements for Upper Scorpius objects have a mean of 3.87, which is very close to the mean of log(g)=3.91 for 5~Myr objects between 2500~and~3000~K predicted by the DUSTY00 evolutionary models. As shown in Figure~\ref{evotracks2}, the distribution of measured surface gravities is only slightly wider than predicted by the DUSTY00 evolutionary models, with outlying values for USco~66AB (log(g)=4.26) and DENIS~1619$-$24 (log(g)=3.49). The relatively high $v$sin$i$ of USco~66AB or higher-order unresolved binarity may contribute to the high surface gravity. Fits to spectra of DENIS~1619$-$24, on the other hand, result in a low surface gravity despite the high $v$sin$i$ and possible binarity. There is no straightforward explanation for the outliers, but they are consistent with the DUSTY00 evolutionary models within the systematic uncertainty in log(g), $\sim$0.5~dex (see below).

 The surface gravity of field dwarfs as measured by fits to the high resolution spectra are somewhat higher than allowed by the DUSTY00 evolutionary tracks, but the measurements also have higher uncertainty because of the increasing mismatch between observed and model spectra for later spectral types and cooler temperatures, caused in large part by the atmosphere models lacking FeH opacity. The surface gravities measured from the N3 spectra are even higher, again suggesting that high-resolution spectra are required to accurately measure surface gravity. It should be noted, though, that our adopted gravities are the weighted mean of results from individual orders that are often either much lower or much higher than what is predicted by the DUSTY00 evolutionary models. 

The age sequence of M6 objects (Table~\ref{tbl-M6age-logg}) generally shows the predicted trend of increasing surface gravity with age, although it is not strictly monotonic. The surface gravities are within $\sim$0.2~dex of what is predicted by the isochrones for their age and effective temperature. A notable exception is 2MASS~2234+40AB, which has the highest effective temperature and surface gravity of the young ($<$100~Myr) objects. However, the adopted temperature and gravity are driven upwards by the poor fits from orders 62 and 63, and temperature and gravity are degenerate such that lowering them both still produces a good spectral fit at high resolution. 

Uncertainty resulting from noise in the observed spectra is determined via the standard deviation of the best-fit surface gravity from flux-re-sampled spectra, which is typically $\lesssim$~0.1~dex but as high as 0.16~dex. As with effective temperature, systematic uncertainty in the accuracy of the atmosphere models likely dominates the errors in the results. The standard deviation of best-fit surface gravities from the varied initial parameters and the standard deviation of the best-ft surface gravities from all orders were typically 0.4--0.6~dex. Systematic uncertainties in surface gravity from the medium-resolution spectra were on average $\sim$0.4~dex. The implications of the relative uncertainties from high- and medium-resolution spectra are discussed in \S~\ref{diss1}.

\subsubsection{Projected Rotational and Absolute Radial Velocities}
\label{sVel}

Results for radial velocity and projected rotational velocity (RV and $v$sin$i$) are presented in Table~\ref{tbl-vel}. We use the well-measured radial velocity of Gl~406 (19$\pm$1~km~s$^{-1}$) to establish systematic uncertainties in our measurements. The radial velocity of GL~406 measured via spectral fitting is 18.7~km~s$^{-1}$, which is consistent with the previously measured RV. Uncertainties are determined using the Monte Carlo implementation of MPFIT (see \S~\ref{fitting}) by taking the weighted mean of standard deviations of the re-sampled-flux results. Typical uncertainties are 1--2~km~s$^{-1}$.

Radial and projected rotational velocities are measured for nine objects for the first time: five Upper Scorpius members, the Taurus object CFHT~Tau~7, the young objects 2MASS~0608$-$27 and Gl~577BC, and the field dwarf LP~402-58.
The average RV of the eight Upper Scorpius members in our sample is $-$7~km~s$^{-1}$ with a dispersion of $<$2~km~s$^{-1}$, consistent with the values of $-$5~km~s$^{-1}$ and $-$6~km~s$^{-1}$ determined by \citet{Muzerolle03} and \citet{Kurosawa06} for five and thirteen very-low-mass Upper Scorpius members, respectively.
The measured radial velocity of SCH~1622$-$19 is furthest from the average of Upper Scorpius members, but still consistent with cluster membership considering our estimated uncertainty of 1--2~km~s$^{-1}$. The measured radial velocity for CFHT~Tau~7 is the peak of the histogram of Taurus members compiled by \citet{Bertout06}, providing more evidence of Taurus membership for this recently-discovered object. RV results for TW Hydrae members are self-consistent with a mean of 8~km~s$^{-1}$ and a standard deviation of $<$2~km~s$^{-1}$; however, they are systematically lower than previous results by 1--6~km~s$^{-1}$.
The RV measurements from our analysis are consistent with other confirmed TW~Hydrae members \citep[e.g.][]{Torres03}. 
The measured RV~=~$-$6.2~km~s$^{-1}$ for Gl~577BC is very similar to the most recently measured RV of the primary, $-$6.5~km~s$^{-1}$ \citep{Nordstrom04}.

In order to test the results from the fitting procedure in which RV is one of four free parameters, we also measure RV by cross-correlating observed spectra with synthetic spectra calculated using the other best-fit parameters (T$_{eff}$, log(g), and $v$sin$i$). The differences were typically less than 1~km~s$^{-1}$ for individual orders, but up to several km~s$^{-1}$ for order 63, particularly for field and late-type objects. The average RV indicated the cross-correlation peak for each order was within 1--2~km~s$^{-1}$ of the weighted mean RV from spectral fitting for all objects, indicating that precision is not sacrificed by including RV as a free parameter in the spectral fitting routine. 

The derived $v$sin$i$ values presented in Table~\ref{tbl-vel} are either consistent with or higher than previously measured values. The high $v$sin$i$ values for slow rotators are an unavoidable limitation of the instrumental resolution, which can be estimated using the known slowly-rotating standard Gl~406 ($v$sin$i$~$<$~2.9~km~s$^{-1}$). Our measured $v$sin$i$ of Gl~406 is 8~km~s$^{-1}$, indicating the lower limit of measurable $v$sin$i$ for the instrumental resolution. Two other objects have similar measured $v$sin$i$ values (DENIS~1605$-$24 and SCH~1612$-$20); therefore, these are considered upper limits on their $v$sin$i$. For the remaining objects the measured $v$sin$i$ is similar to or slightly higher than $v$sin$i$ values reported in the literature with the largest discrepancy for the field object LP~412-31. Both LP~412-31 and 2MASS~0140+27 are poorly reproduced by even the best-fit model parameters because of the lack of FeH opacity in the atmosphere models, which become a major source of spectral features for late M dwarfs. When the model spectrum does not match the observations well enough to constrain $v$sin$i$, the fitting routine tends to return a high value of $v$sin$i$ because it flattens out the model spectrum to compensate for mismatched features, thus minimizing the $\chi^2$. However, it is unclear why the best-fit for on object would produce a much higher $v$sin$i$ when the previously published values are at or below the instrumental resolution, as is the case for LP~412-31. LP~412-31 is a strongly flaring M~dwarf with very strong H$\alpha$ emission and magnetic field for its spectral type \citep{Reid02,Stelzer06,Schmidt07,Reiners07b}; perhaps the high $v$sin$i$ measurement in the near-infrared is related to activity (the previously published $v$sin$i$ was measured in the optical by \citealt{Reid02}). The estimated precision of the $v$sin$i$ measurements is $\pm$5~km~s$^{-1}$ based on the typical standard deviation of measurements from each order.

\section{Discussion}

\subsection{Measuring the Physical Properties of Young Brown Dwarfs}
\label{diss1}
Model fits to medium-resolution spectra produce consistent estimates of effective temperature but only weakly constrain surface gravity. On the other hand, high-resolution fits can be misleading because of the degeneracy between effective temperature, surface gravity, and $v$sin$i$, which all contribute to broadening of the strongest atomic lines. Weaker molecular lines can help break this degeneracy, but the imperfect correspondence between observations and atmosphere models for the molecular lines, particularly FeH, confuse this issue. Therefore, the physical properties of brown dwarfs are likely best estimated by a combining both medium- and high-resolution for simultaneous fitting.

Representative observed and best-fit model spectra are presented in Figures~\ref{DE1605fit}--\ref{l40258fit}. From a comparison of the best-fit spectra for the young (5~Myr) M6 object DENIS~1605$-$24 (Fig.~\ref{DE1605fit}) and the field ($>$~1~Gyr) M6 dwarf Gl~406 (Fig.~\ref{gl-406fit}), it is evident that both the atomic and molecular lines are stronger in the field object that in the young object. The best-fit synthetic spectra are a better match for the young object than for the field object, particularly in the wings of the atomic lines. As described in \S~\ref{bdmodels}, this is possibly a result of the simplified line profiles that are currently implemented in the {\tt PHOENIX} code. However, it is likely that shape of the line wings are also affected by blended molecular lines, especially FeH, that are not yet properly reproduced by the model (see \S~\ref{sFeH}). It is promising, however, that in the spectra of DENIS~1605$-$24 there is a substantial correspondence between weak molecular lines in the model, especially evident in orders 63--61. The same lines are present in the best-fit spectra for Gl~406, but the line depths are far too shallow in the model even for this relatively hot object. Note that the missing molecular opacity in the model is also apparent in the medium-resolution spectrum of Gl~406 (last panel, gray) compared to the model spectrum with the weighted best-fit parameters from the high-resolution spectra. However, individual molecular lines can only be examined at high resolution. 

Figures~\ref{2m1139fit} and ~\ref{l40258fit} are representative of the fits for later spectral type objects at different ages ($\sim$10~Myr: 2MASS~1139$-$31 and $>$1~Gyr: LP~402-58). For 2MASS~1139$-$31 there are considerably more mismatches in the best-fit spectra than there are for the slightly younger and earlier-type object DENIS~1605$-$24. While the strong K~{\sc i} lines and many of the weaker features in the order 65 and 61 of 2MASS~1139$-$31 are reproduced, the strongest lines in orders 64, 59, and 58 are too deep in the model spectra. This can be attributed to the missing FeH opacity that dominates the molecular lines at these wavelengths: the pseudo-continuum level in the synthetic spectra is too high because of the missing opacity, resulting in atomic lines that are too deep when the synthetic pseudo-continuum is lowered to the observed pseudo-continuum. In the order 62 and 63 spectra the increasingly poor reproduction of the FeH bands at lower temperature and higher surface gravity is evident, although some features are coincident in wavelength. The best-fit parameters also reproduce the overall continuum shape of the medium-resolution spectrum of 2MASS~1139$-$31 (last panel of Fig.~\ref{2m1139fit}), although there is missing opacity from 1.19--1.24~$\mu$m, similar to Gl~406 in Fig.~\ref{gl-406fit}. The sharp drop in flux at 1.24~$\mu$m is less pronounced when a more detailed calculation of atomic line profiles are implemented in the code (see~\ref{bdmodels}). The spectral fit for field M7 dwarf LP~402-58 (Fig.~\ref{l40258fit}) also shows the success of the fit for the K~{\sc i} lines in order 65 and 61 and the too-deep Fe~{\sc i}, Mg~{\sc i}, and Al~{\sc} lines in order 64, 59, and 58. Similarly, the FeH bands in orders 63 and 62 are too weak in the model, likely contributing to the mismatches in atomic line depths at shorter wavelengths as well. The pseudo-continuum slope of the medium-resolution spectrum is flatter in the observed spectrum than in the model, suggesting that a cooler effective temperature provides a better fit. An effective temperature 100--200~K lower flattens out the pseudo-continuum but also produces a too-deep H$_2$O band at 1.34~$\mu$m and atomic lines that are too broad at high resolution, even in combination with lower surface gravity and $v$sin$i$ values. Improving the FeH opacity in the atmosphere models is a likely first step to resolving this issue.

Comparing our adopted values of T$_{eff}$ and log(g) to isochrones of the DUSTY00 evolutionary models \citep{Chabrier00,Baraffe02} provides a consistency check and a means of evaluating our results. Even though the DUSTY00 evolutionary models are poorly calibrated for the youngest objects, they can rule out unphysical values. For example, objects 5~Myr old with temperatures from 1800~K to 3000~K all have log surface gravities between 3.81 and 3.96, objects older than 500~Myr in the same temperature range have log surface gravities between 5.15 and 5.40, and no object of any age or effective temperature has a surface gravity over log(g)=5.40. 

All of our adopted temperatures and surface gravities are consistent with predictions of the DUSTY00 evolutionary models. Evolutionary tracks can be used to infer the masses and ages of objects from their measured temperatures and surface gravities. Adopted T$_{eff}$ and log(g) values for the 5--10~Myr subsample (including 2MASS~0608$-$27) and additional objects with ages~$\le$1--3~Myr are compared to the DUSTY00 tracks in Figure~\ref{evotracks2}. Within the error bars all of the objects are consistent with the properties predicted for young (age~$<$~20~Myr) brown dwarfs (mass~$<$~72~M$_J$). Two slightly anomalous objects are DENIS~1619$-$24 (too low gravity) and USco~66AB (too high gravity), which are discussed above. All of the objects are well above the deuterium-burning mass limit \citep[$\sim$13~$M_J$,][]{Chabrier00}. 

The relative systematic uncertainties on the best-fit effective temperatures and surface gravities from high- and medium-resolution spectra indicate a promising method of inferring the physical properties of young brown dwarfs. Medium-resolution $J$-band spectra strongly constrain effective temperature of mid-late M dwarfs, but not surface gravity. Neither can medium-resolution spectra be used to measure $v$sin$i$, which may influence surface gravity measurements. High-resolution spectra, particularly of the shorter-wavelength K {\sc i} doublet in NIRSPEC dispersion order 65, strongly constrain surface gravity relative to effective temperature. Therefore, we anticipate that using a measurement of effective temperature from medium-resolution spectra to constrain the effective temperature for high-resolution spectral fitting will increase the accuracy of properties derived using high-resolution spectra. The measured range of $v$sin$i$ values does not correlate with age or known binarity. Improvements in the model atmospheres that more consistently reproduce the spectral features of both young and field brown dwarfs will improve the accuracy of $v$sin$i$ measurements and provide insight into the dependence of angular momentum evolution on mass and age.

\subsection{FeH Oscillator Strengths}
\label{sFeH}
One of the glaring mismatches between observed and synthetic spectra is the apparently missing opacity from FeH bands, particularly in orders 62 and 63. Our model atmosphere calculations incorporate new FeH line list from \citet{Dulick03} as an update to the previously used \citet{Phillips87} line list. However, one key difference between the use of these line lists in the {\tt PHOENIX} code is that oscillator strength scalings available for the \citet{Phillips87} line list are not applied to the \citet{Dulick03} line list. \citet{Burgasser03} shows that the {\it absorption coefficients} in the \citet{Dulick03} line list correspond to absorption features in the medium-resolution $J$-band spectrum of a late-L dwarf (their Figure~2b). Our results indicate that the \citet{Dulick03} line list, as incorporated into the {\tt PHOENIX} code, fails to reproduce observed FeH absorption features.

Figures~\ref{fehmed} and \ref{fehhi} illustrate the effects of different FeH line lists and oscillator strength scalings on synthetic spectra. Figure~\ref{fehmed} shows $J$-band spectra synthetic spectra (T$_{eff}$=2600~K, log(g)=5.0, convolved to match a medium-resolution NIRSPEC spectrum) for the \citet{Dulick03} FeH line list, the \citet{Phillips87} line list, and the ``scaled'' \citet{Phillips87} line list, for which red-optical and near-infrared FeH band strengths are multiplied by factors of 10-20 (and for one band, 150). It is evident from comparison with the observed spectrum of 2MASS~0140+17 (M9) that the scaled FeH strengths using the \citet{Phillips87} line lists match the observed spectra better, particularly the broad absorption features at 1.20 and 1.21 $\mu$m that are essentially absent in spectra with unscaled line lists. At this resolution the difference between the unscaled line lists is imperceptible, but the high-resolution spectra provide further insight. In Figure~\ref{fehhi} the same synthetic spectra are convolved to match NIRSPEC dispersion order~63, and it is apparent that even the scaled \citet{Phillips87} line list falls short of reproducing the observed spectrum of the M9 dwarf 2MASS~0140+27. The \citet{Dulick03} list provides more individual lines in this wavelength range, but the lines are still far too weak. In a future paper we will quantify the shortcomings in the current FeH data by implementing scaling of the \citet{Dulick03} line list in the {\tt PHOENIX} code.

FeH is an important source of spectral features from the $z$ band through the $H$ band, particularly in the longer wavelength half of the $J$ band \citep{McLean03,Cushing03,Cushing05,McLean07}, and the features strengthen with decreasing effective temperature until $\sim$L5 \citep{Kirkpatrick99}, when FeH begins to weaken through the L-T transition then strengthen in mid-to-late T dwarfs, possibly as cloud-clearing allows flux to escape from lower in the atmosphere where FeH remains as a gas \citep{Burgasser02,McLean03,Cushing08}. Therefore, further study of FeH in the atmosphere models is necessary to extend our near-infrared spectral fitting analysis to L and T dwarfs as well as to shorter and longer wavelength observations.

Magnetic fields may also affect FeH strengths, and the Wing-Ford band of FeH ($\sim$1~$\mu$m) has been used to measure magnetic field strengths in M dwarfs. However, strong magnetic fields actually weaken the magnetically-sensitive lines \citep[see][Figures~2--10]{Reiners07b}. Therefore, if magnetic fields need to be included in the atmosphere models in order to reproduce observed FeH bands, it is likely a second-order effect to the overall scaling of the oscillator strengths.

\subsection{Effects of Binarity}
\label{binarity}
Unresolved binaries in the sample are Gl~577BC, USco~66AB, 2MASS~2234+40AB, and 2MASS~1207$-$39Ab. These binaries vary in mass ratio ($q$) from $q\sim$1 (Gl~577BC: \citealt{Lowrance05}, USco~66AB: \citealt{Kraus05}, 2MASS~2234+40AB: \citealt{Allers09}) to $q\sim$0.3 (2MASS~1207$-$39Ab: \citealt{Mohanty07}). Additionally, DENIS~1619$-$24 is a candidate spectroscopic binary \citep{Mohanty05}, and \citet{Slesnick06} suggests that SCH~1622$-$19 might be an unresolved binary (a~$\lesssim$~175~AU) based on overluminosity for its spectral type relative to other Upper Scorpius members on an H-R diagram. For our purposes the effects of unresolved binarity are minimal as long as the mass ratio is either very high or very low. The main concern is the effect of unresolved binarity on the estimate of $v$sin$i$. Previous authors have noted that broadened absorption lines may be indicative of an unresolved binary \citep[e.g.][]{Simon06}. As shown in Figure~\ref{binarytest}, the velocity shift between components of a close binary can mimic a higher value of $v$sin$i$. However, the RV-shift needed to reproduce the spectrum of USco~66AB ($v$sin$i$~=~28 km~s$^{-1}$), using the DENIS~1605$-$24 ($v$sin$i$~$\le$~7 km~s$^{-1}$) as a template, is at least several times larger than the maximum possible velocity difference of the known binary. Two possibly unresolved binaries, DENIS~1619$-$24 and SCH~1622$-$19, are among the highest $v$sin$i$ measurements, but they are not anomalous compared to other, apparently single, rapidly rotating objects. A future paper will explore this issue further by comparing spectral fitting results from spatially resolved (using NIRSPEC and Laser Guide Star Adaptive Optics, e.g. Konopacky et al. in prep.) and unresolved spectra of very low mass binaries.

\section{\sc Conclusions}
We have estimated the effective temperature, surface gravity, absolute
radial velocity and projected rotational velocity for a sample of
young brown dwarfs and field M dwarfs. Comparison of high-resolution
synthetic and observed spectra of young objects provide more accurate determination of
surface gravity than from lower resolution observations. However, the small wavelength range and
lack of fidelity of FeH line strength in the atmosphere models result in
anomalously high temperatures for late-M type objects. Despite the mismatch in line strengths,
radial velocity is well-determined for all but the coolest field
objects considered in this analysis. Additionally, the richness of
lines at high resolution breaks the degeneracy between sources of
line-broadening so that $v$sin$i$ is also well-determined for most
objects. The principal results of the paper are as follows:

(1) Adopted (weighted-mean) effective temperatures from high-resolution spectral fits for objects M7 and earlier fall within or very close to previously published estimates and results from medium-resolution spectral fits. For objects M8 and later, temperatures from medium-resolution fits are similar to previously published values and temperatures from high-resolution fits are systematically too hot.

(2) Adopted (weighted-mean) surface gravities from high-resolution fits neatly separate objects with T$_{eff}$ from 1800--3000~K according to age, with 1--10~Myr objects lying broadly between log(g)=3.5 and 4.3, while objects from 0.5--5~Gyr are in the log(g)=5.2--5.4 range. Surface gravities from medium-resolution fits are generally similar to the adopted values except for M8--M9 young objects, which are systematically low, and field objects, which are systematically high.

(3) The age sequence of M6 objects yields an average temperature of 2880~$\pm$~60~K.

(4) For the Upper Scorpius objects the mean surface gravity is log(g)=3.87, which is very close to the value of log(g)=3.91 predicted by the DUSTY00 evolutionary model for 5~Myr objects between 2500 and 3000~K.

(5) FeH features are not yet properly reproduced by atmosphere models, and this impacts fitting spectra to high-resolution observations. The recent \citet{Dulick03} line lists provide better correspondence for individual features, but the lines are not strong enough in the models. Improving FeH in the atmosphere models is urgently needed.

(6) Physical properties of brown dwarfs are likely best measured by comparing observed and synthetic spectra using a combination of medium- and high-resolution spectra simultaneously.

This work represents the first in a series of analysis of medium- and high-resolution observed and synthetic spectra of brown dwarfs. Future work will extend the study to later spectral types, cooler atmosphere models (where T$_{eff}$ and log(g) might become more intertwined, e.g. \citealt{Burgasser06}), and different dust treatments (i.e. the {\tt PHOENIX}-{\it cond} models). Further analysis will also explore effects of metallicity on atmospheric structure and resultant spectra. While the focus thus far is on the large database of high-resolution spectra provided by the BDSS, some combination of high resolution and broad spectral coverage will likely prove ideal for accurately inferring the physical properties of brown dwarfs using atmosphere models. Therefore, we are developing a procedure to combine all available observed spectral data for simultaneous model fits.

\acknowledgements
The authors wish to thank the staff of the Keck Observatory for their outstanding support, including Joel Aycock, Randy Campbell, Al Conrad, Grant Hill, Jim Lyke, Steven Magee, Julie Renaud-Kim, Barbara Schaefer, Chuck Sorenson, Terry Stickel, and Cynthia Wilburn. Observing assistance from Antonia Hubbard, Quinn M. Konopacky, Gregory Mace, and Erin C. Smith and data reduction by Chalence Safranek-Schrader was greatly appreciated. E.L.R. acknowledges the hospitality of the research and administrative staff at Lowell Observatory. T.B. acknowledges the NASA Origins of Solar System program and the Mount Cuba Astronomical Fund for their generous support. I.S.M. acknowledges the staff of the UCLA Infrared Laboratory and colleagues James Graham (UCB), James Larkin (UCLA) and Eric Becklin (UCLA) for their support throughout the development of the NIRSPEC instrument. This paper benefited greatly from the detailed and thoughtful comments of the anonymous referee.

This research has made use of the NASA/IPAC Infrared Science Archive, which is operated by the Jet Propulsion Laboratory, California Institute of Technology, under contract with the National Aeronautics and Space Administration. This publication makes use of data from the Two Micron All Sky Survey, which is a joint project of the University of Massachusetts and the Infrared Processing and Analysis Center, funded by the National Aeronautics and Space Administration and the National Science Foundation. This research has benefited from the M, L, and T dwarf compendium housed at DwarfArchives.org and maintained by Chris Gelino, Davy Kirkpatrick, and Adam Burgasser. This research has made use of the SIMBAD database, operated at CDS, Strasbourg, France and NASA's Astrophysics Data System. Finally, the authors wish to extend special thanks to those of Hawaiian ancestry on whose sacred mountain we are privileged to be guests.

{\it Facilities:} \facility{Keck:II (NIRSPEC)}

\bibliography{/Users/erice/Documents/UCLA/THESIS/thesis}

\begin{figure}
  \includegraphics[height=.65\textheight,angle=90]{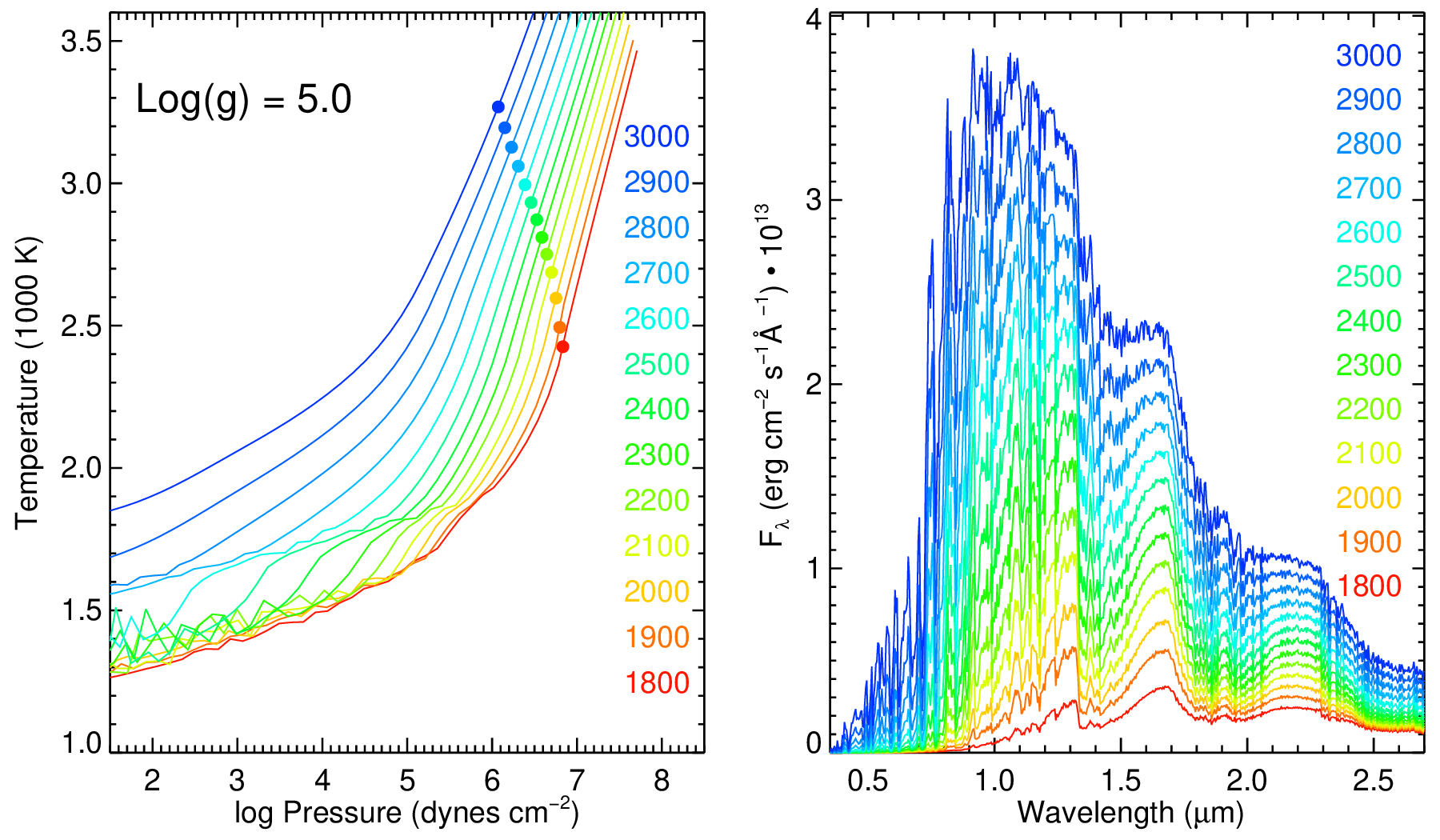}
  \includegraphics[height=.65\textheight,angle=90]{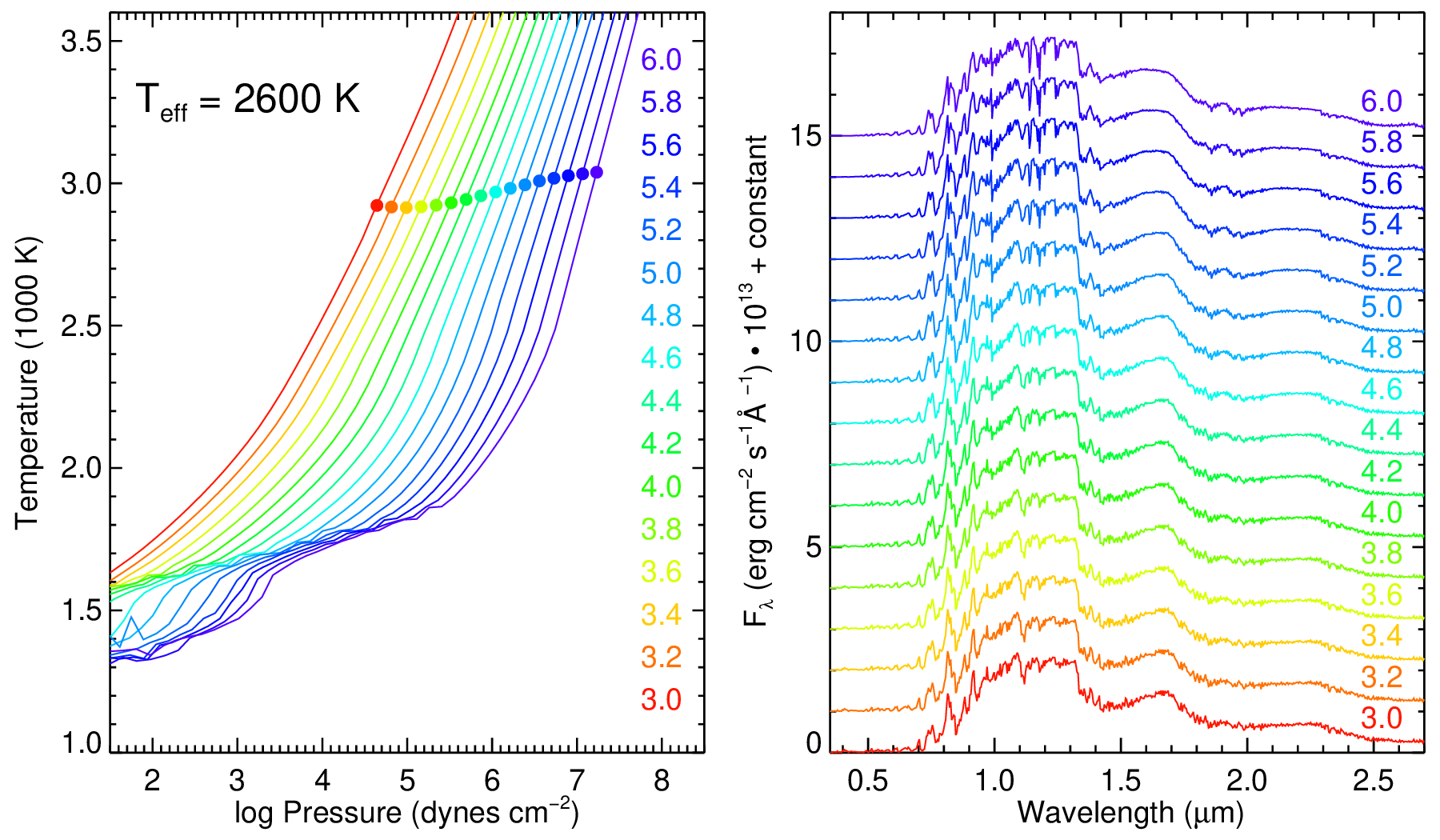}
  \caption{\label{models}Left: Temperature-pressure profile of model atmosphere
    structures calculated using the {\tt PHOENIX} code. Filled circles
    mark the near-infrared photosphere
    ($\tau$$_{\lambda=1.2~\micron}$~=~$\frac{2}{3}$). Right: Spectral
    energy distribution (SED) for the same model atmospheres. The top
    plots display structures and SEDs calculated at log(g)=5.0 for the
    entire range of effective temperatures. The bottom plots display structures and
    SEDs (offset by a constant) calculated at T$_{eff}$=2600~K for the entire range of
    surface gravities. Only half of the calculated structures and SEDs
    are shown for clarity. 
}
\end{figure}

\clearpage

\begin{figure}
  \includegraphics[height=.75\textheight,angle=90]{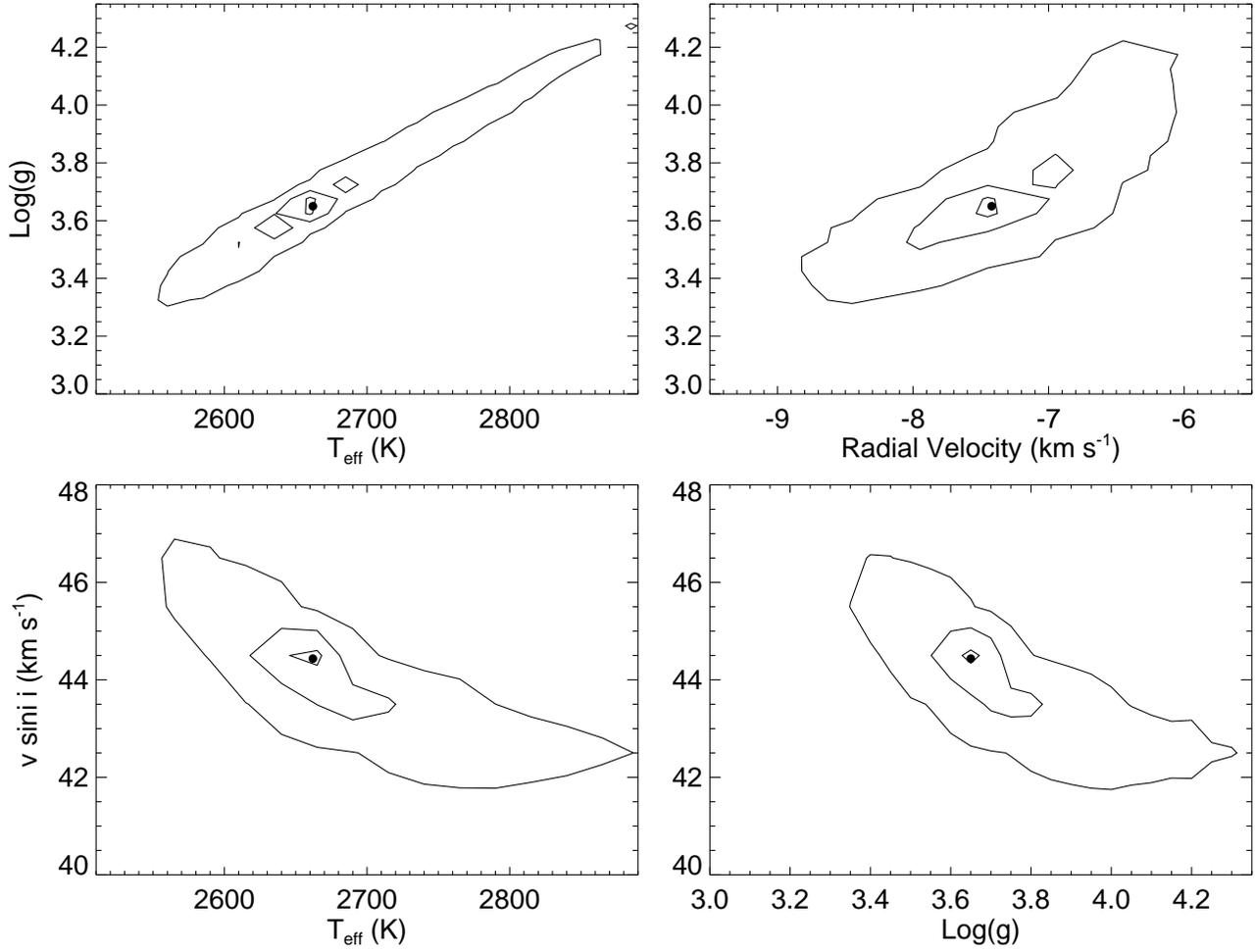}
  \caption{\label{chis} The distribution of best-fit values found by
  the MPFIT procedure with varied input parameters (see \S3.3) for the
  order 65 spectrum (1.165 -- 1.182 $\mu$m) of
  DENIS~1619$-$24. Contours mark 10\%, 50\%, and 90\% of the maximum
  number of points in 25~K, 0.05 dex, 0.5 km~s$s^{-1}$ (RV) and 0.5
  km~s$s^{-1}$ ($v$sin$i$) bins. Filled circles mark our adopted best-fit values. Top
  left: Effective temperature and surface gravity show a strong
  degeneracy in this wavelength range. Top right: Radial velocity is
  well-determined independent of the other parameters.  Bottom:
  Projected rotational velocity is slightly degenerate with both
  effective temperature and surface gravity. These plots do not show
  the $\chi^2$ value for each fit result, which is taken into account
  to determine our adopted best-fit parameters.}
\end{figure}

\clearpage

\begin{figure}
  \includegraphics[height=0.75\textheight,angle=90]{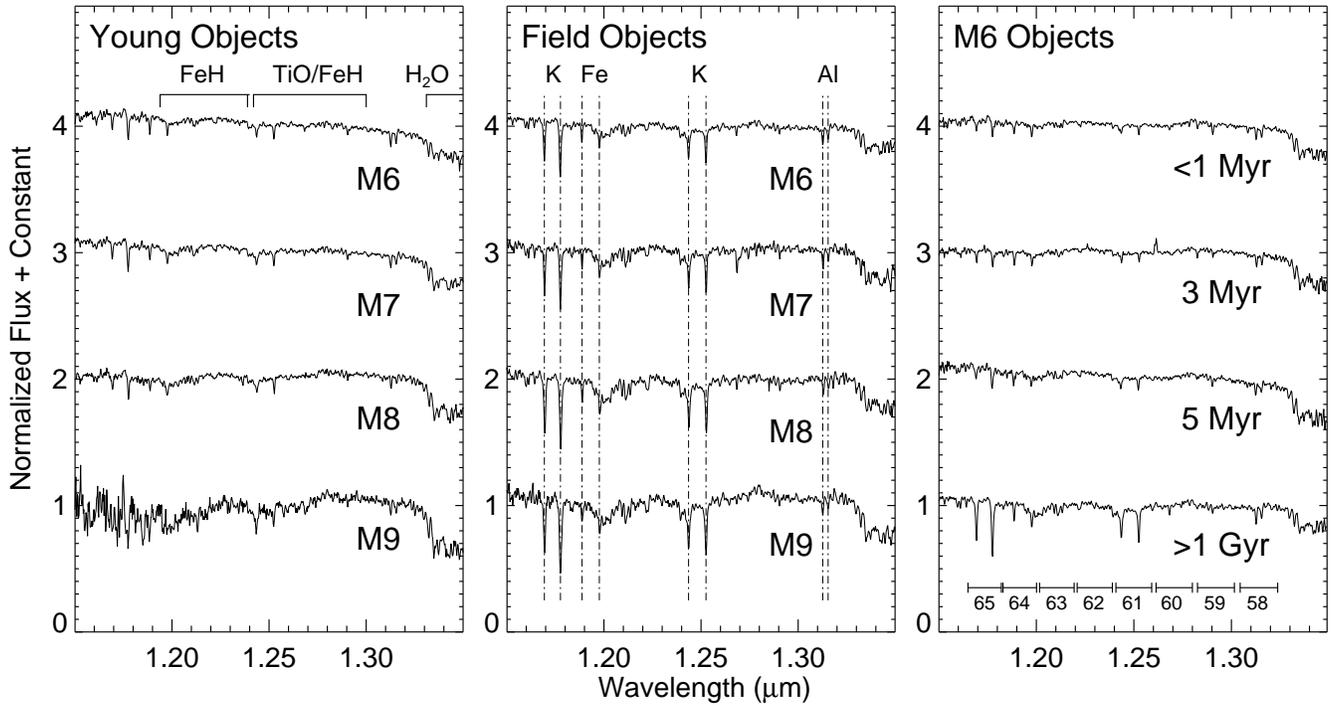}
  \caption{\label{medres3} Observed $J$-band spectra of  young (age
    $\sim$5--10~Myr) M dwarfs (left), field (age $\ge$~1~Gyr) M dwarfs
    (middle), and M6 objects with a range of ages (right). Spectra are
    normalized at 1.26 $\mu$m and offset by a constant. The left plot
    labels molecular absorption bands, the middle plot strong atomic
    lines, and the right plot the wavelength coverage of NIRSPEC
    dispersion orders. 
}
\end{figure}

\clearpage

\begin{figure}
  \includegraphics[height=.80\textheight,angle=90]{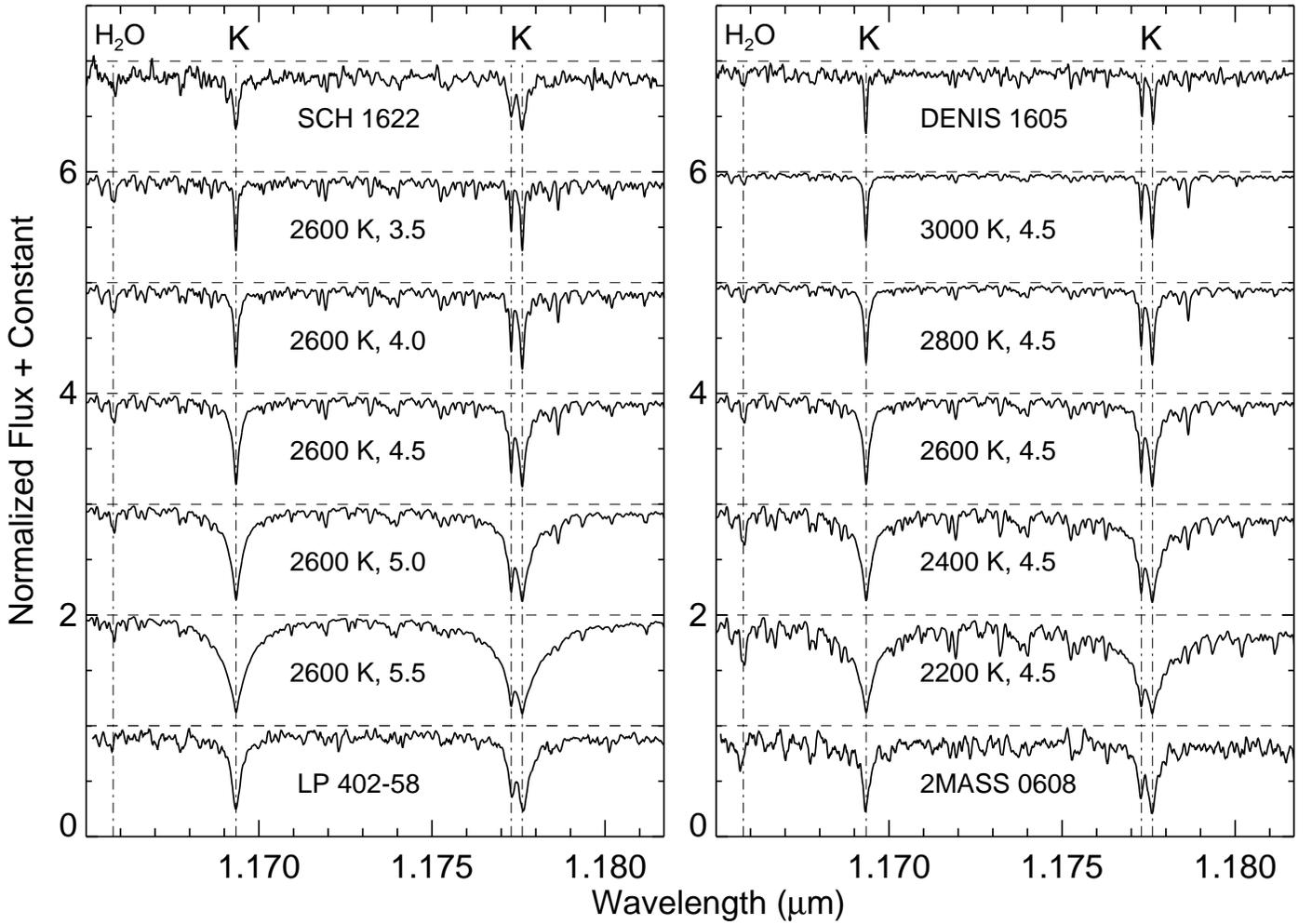}
  \caption{\label{models65}High-resolution (R$\sim$20,000) observed
    and synthetic spectra for NIRSPEC dispersion order 65. In both panels
    at top and bottom are observed spectra (SCH~1622$-$19: M8, 5
    Myr; LP~412-31: M8, $>$1~Gyr; DENIS~1605$-$24: M6, 5~Myr; 2MASS~0608$-$27:
    M8.5, $<$100~Myr) whereas the five middle spectra are synthetic
    spectra with effective temperature and log surface gravity as
    labeled. Horizontal lines denote vertical offsets in the pseudo-continuum levels. 
    The strongest absorption lines are marked by
    dot-dashed lines and labeled on the plot. The
    effects of increasing surface gravity (left) and decreasing
    temperature (right) on the pressure-sensitive K {\sc i} lines are
    similar but can be disentangled at high resolution, e.g. using the
    strength of the H$_2$O line at 1.16579 $\mu$m and other molecular lines.
}
\end{figure}

\clearpage

\begin{figure}
  \includegraphics[height=.80\textheight,angle=90]{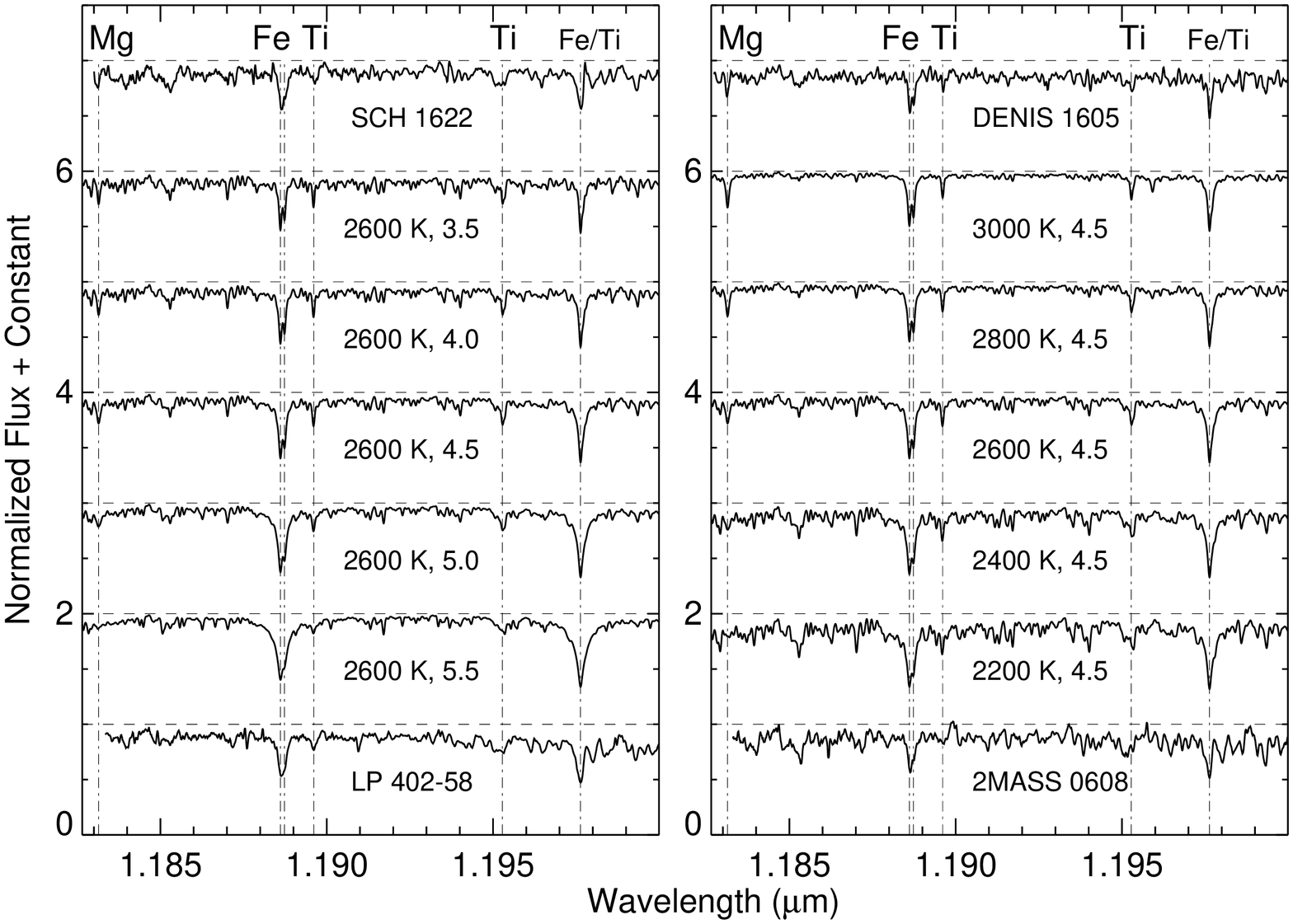}
  \caption{\label{models64}The same as Figure~\ref{models65} but for
    NIRSPEC dispersion order 64. }
\end{figure}

\clearpage

\begin{figure}
  \includegraphics[height=.80\textheight,angle=90]{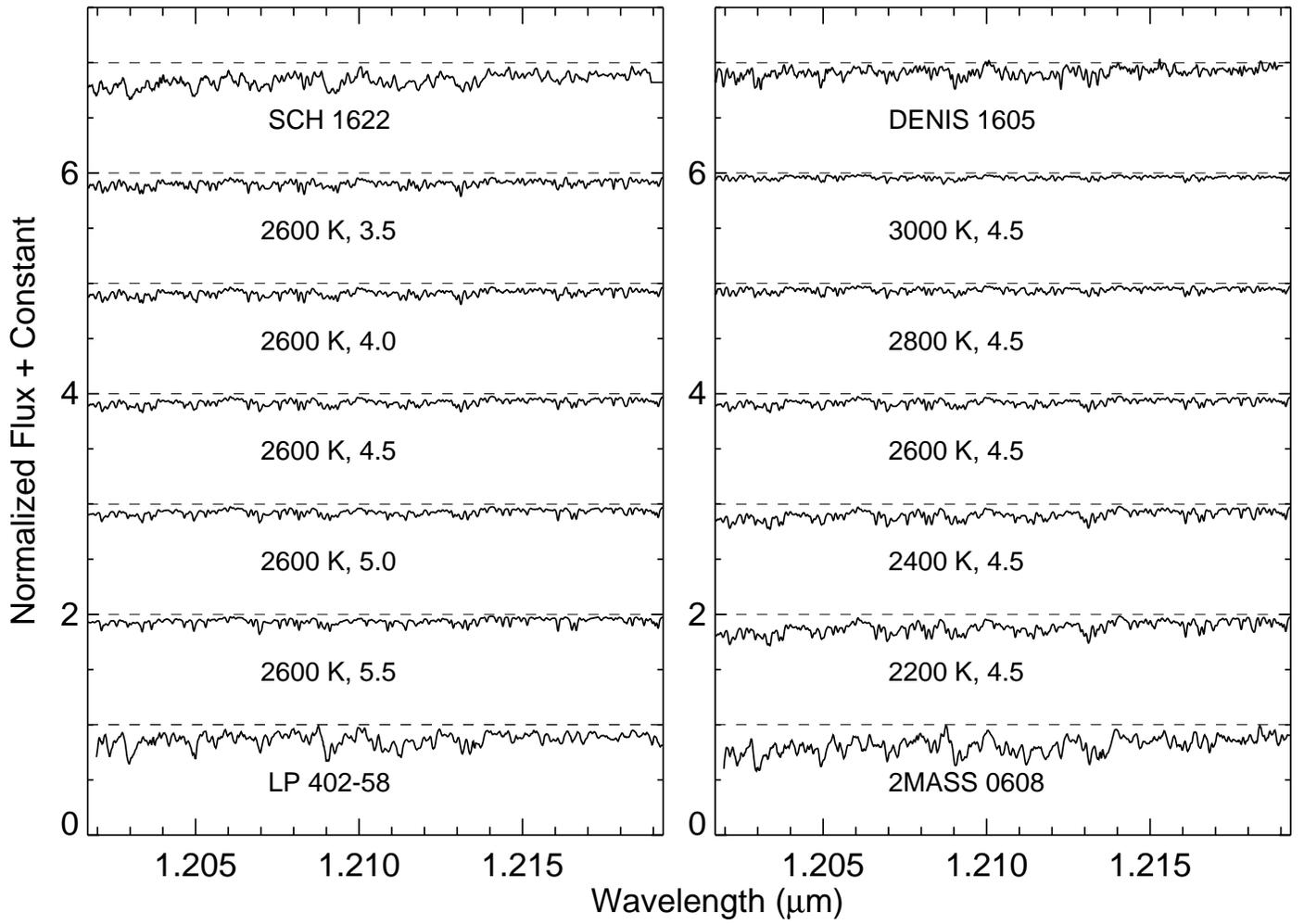}
  \caption{\label{models63}The same as Figure~\ref{models65} but for
    NIRSPEC dispersion order 63. The spectral features in this order
    are primarily H$_2$O and FeH, the latter of which is too weak in
    the atmosphere models (see \S~\ref{sFeH}).
}
\end{figure}

\clearpage

\begin{figure}
  \includegraphics[height=.80\textheight,angle=90]{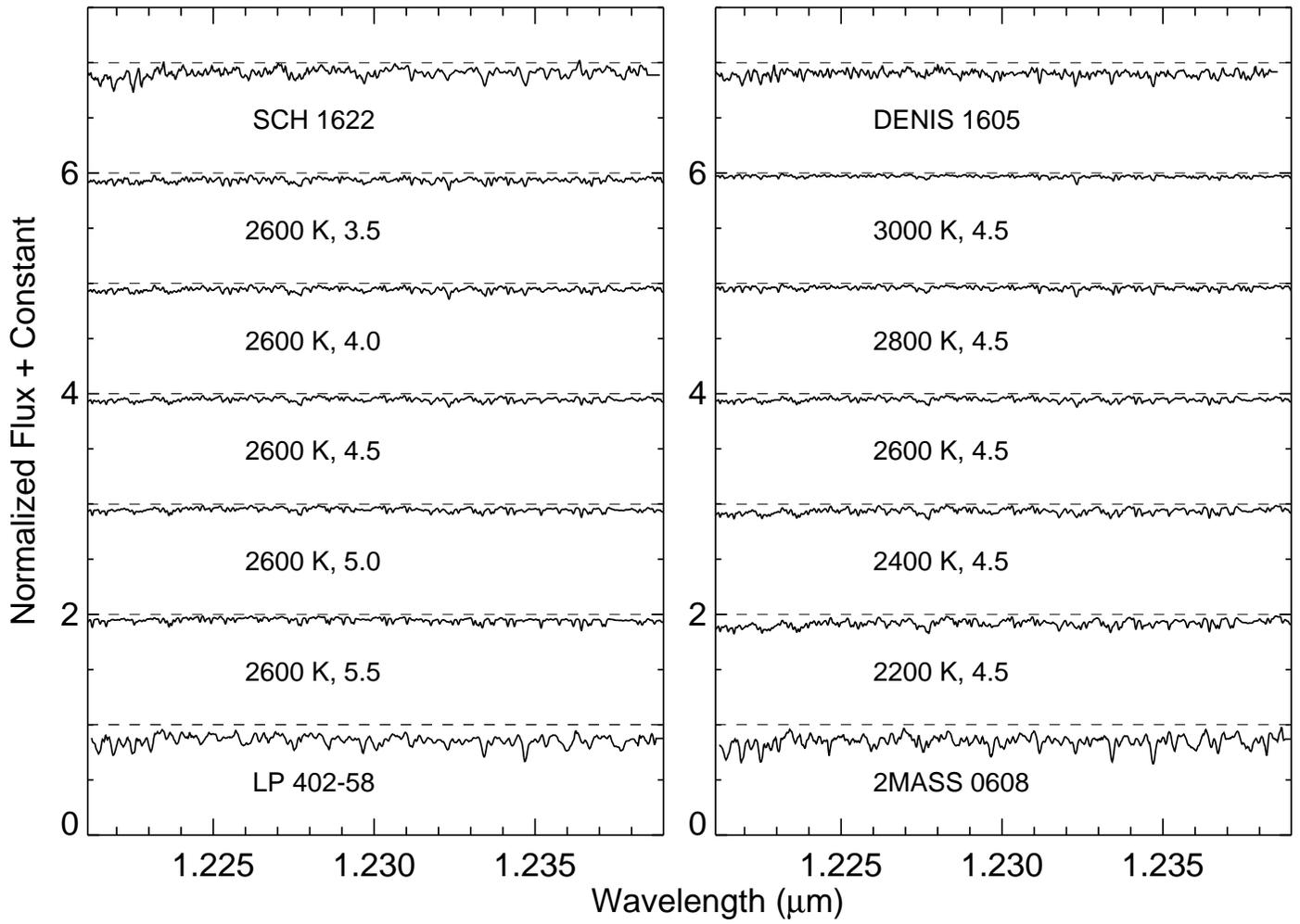}
  \caption{\label{models62}The same as Figure~\ref{models65} but for
    NIRSPEC dispersion order 62. The spectral features in this order
    are primarily H$_2$O and FeH, and again FeH is too weak in the
    atmosphere models (see \S~\ref{sFeH}), notably the the $Q$-branch of the
    $F^{4}$$\Delta_{7/2}$-$X^{4}$$\Delta_{7/2}$ system from 1.221 to 1.223 $\mu$m.
}
\end{figure}

\clearpage

\begin{figure}
  \includegraphics[height=.80\textheight,angle=90]{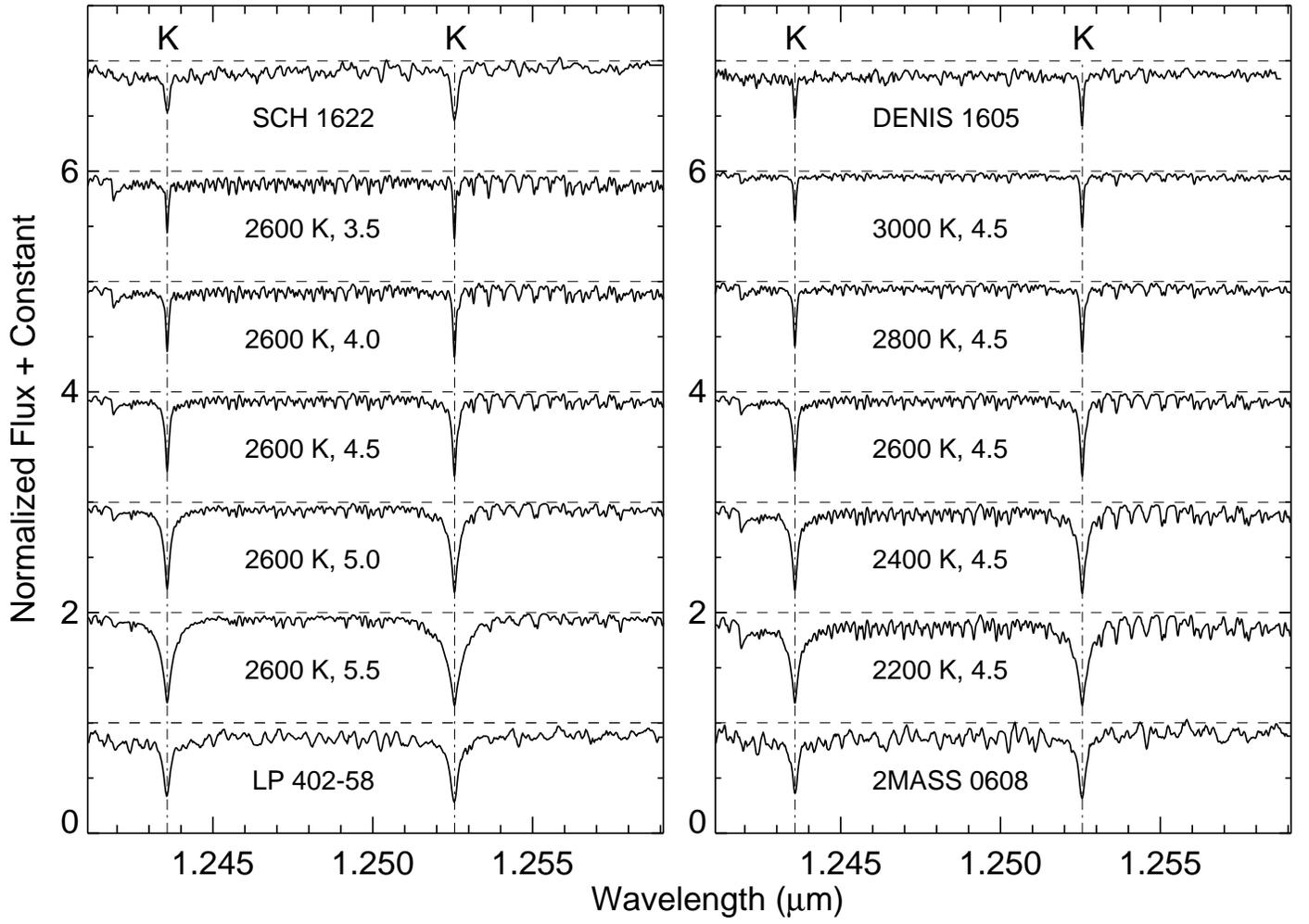}
  \caption{\label{models61}The same as Figure~\ref{models65} but for
    NIRSPEC dispersion order 61. Regularly-spaced lines in the synthetic spectra
    are from TiO. In the observed spectra TiO (if present) is blended with
    FeH, but FeH is considerably weaker in the atmosphere models (see \S~\ref{sFeH}).
}
\end{figure}

\clearpage

\begin{figure}
  \includegraphics[height=.80\textheight,angle=90]{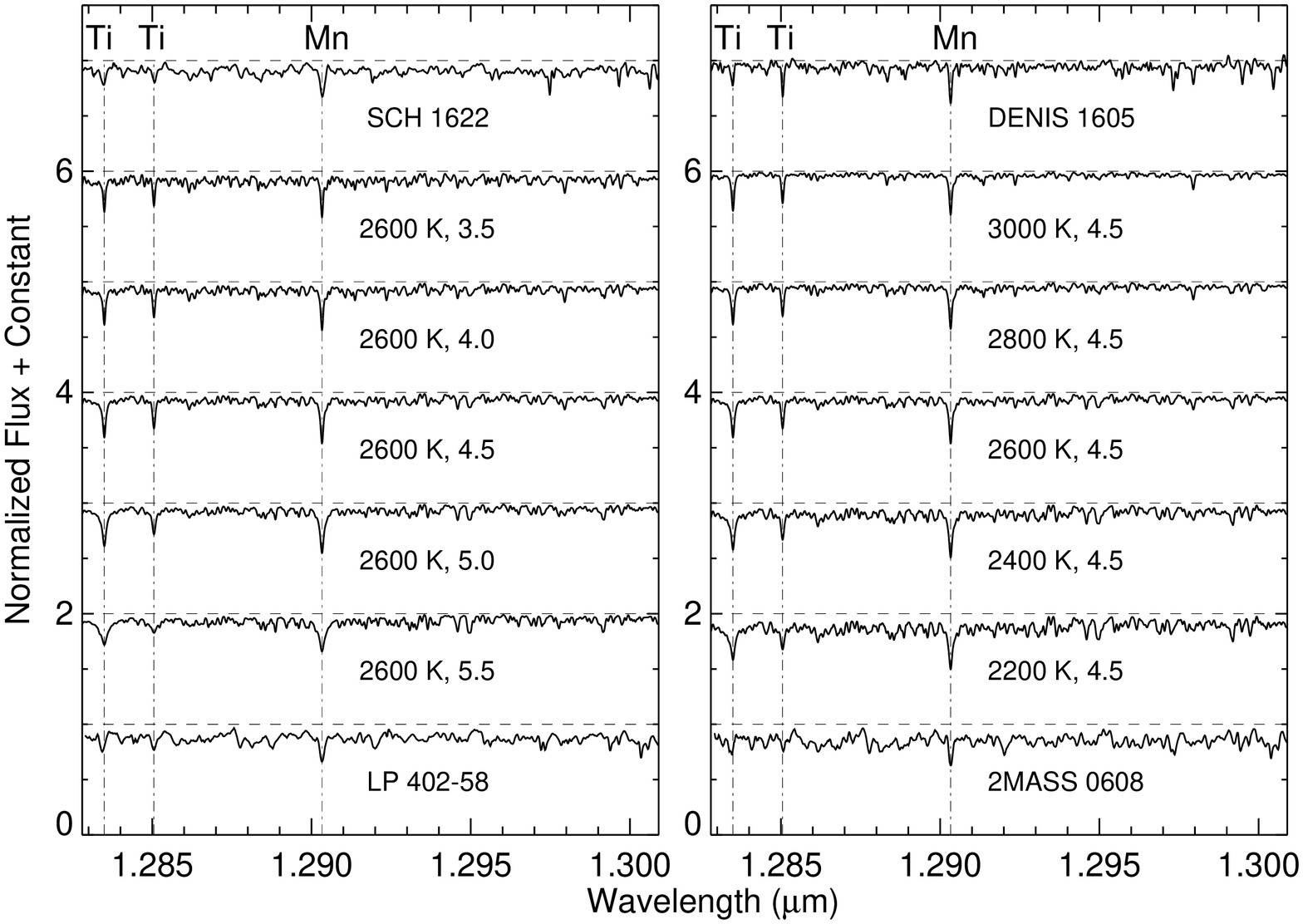}
  \caption{\label{models59}The same as Figure~\ref{models65} but for
    NIRSPEC dispersion order 59. 
}
\end{figure}

\clearpage

\begin{figure}
  \includegraphics[height=.80\textheight,angle=90]{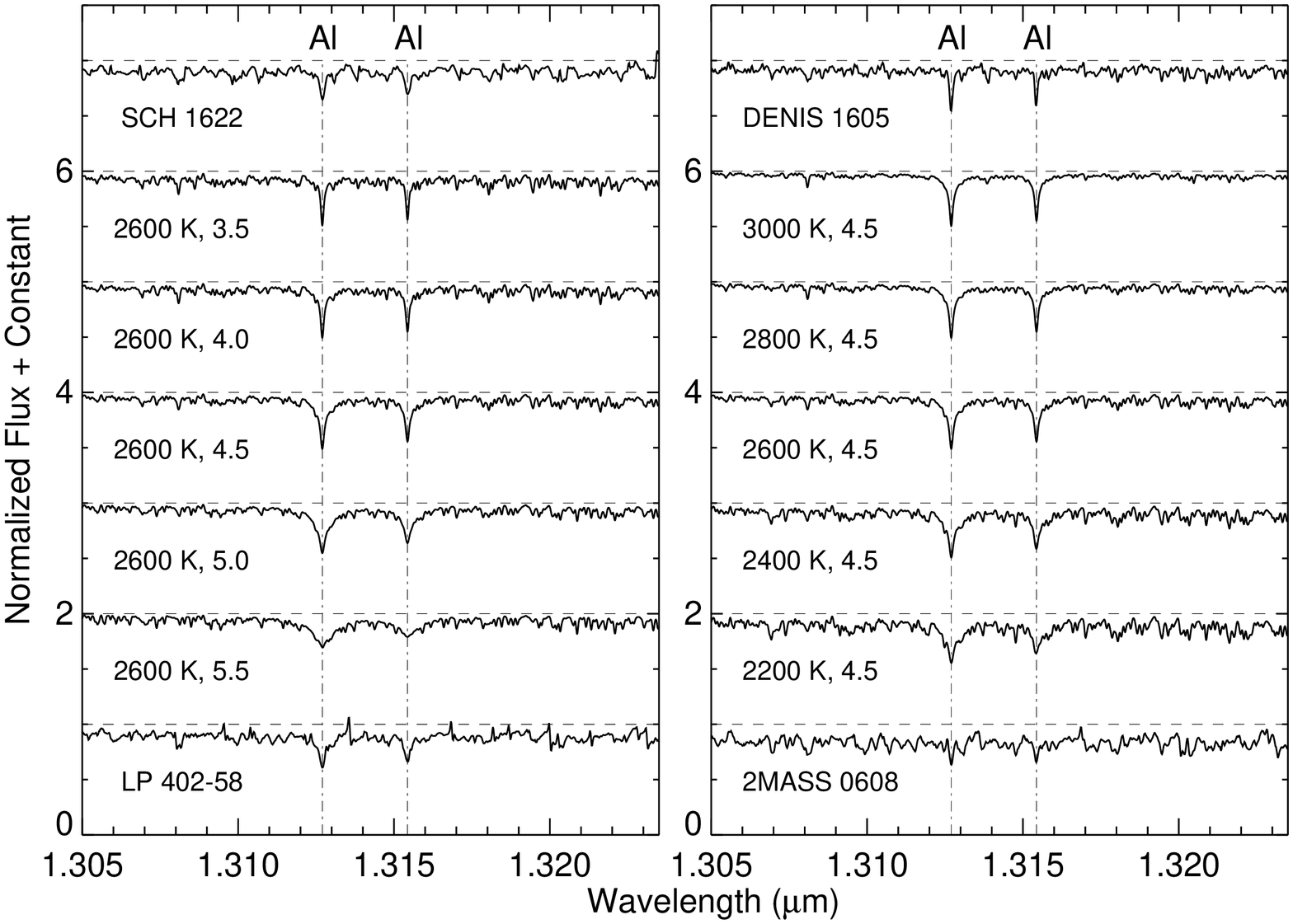}
  \caption{\label{models58}The same as Figure~\ref{models65} but for
    NIRSPEC dispersion order 58.}
\end{figure}

\clearpage

\begin{figure}
  \includegraphics[height=.9\textheight,angle=0]{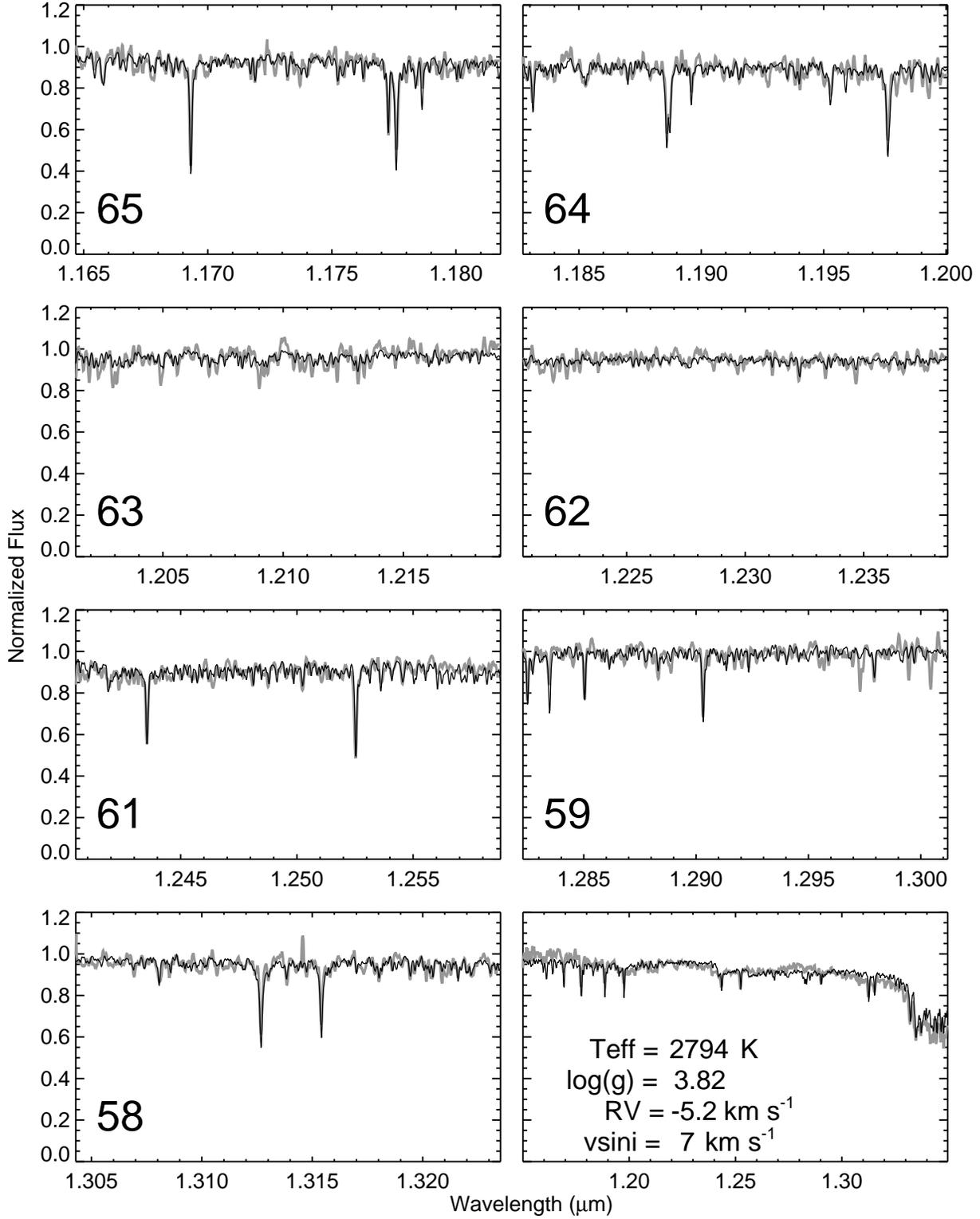}
   \caption{\label{DE1605fit} Observed (gray) and best-fit synthetic
     spectra (black) for the young (5~Myr) M6 object DENIS~1605$-$24. The
     NIRSPEC dispersion orders are labeled, and the best-fit atmosphere model
     parameters are given in the bottom-right panel, which is the
     medium-resolution spectrum across the entire $J$ band.
}
\end{figure}

\clearpage

\begin{figure}
  \includegraphics[height=.9\textheight,angle=0]{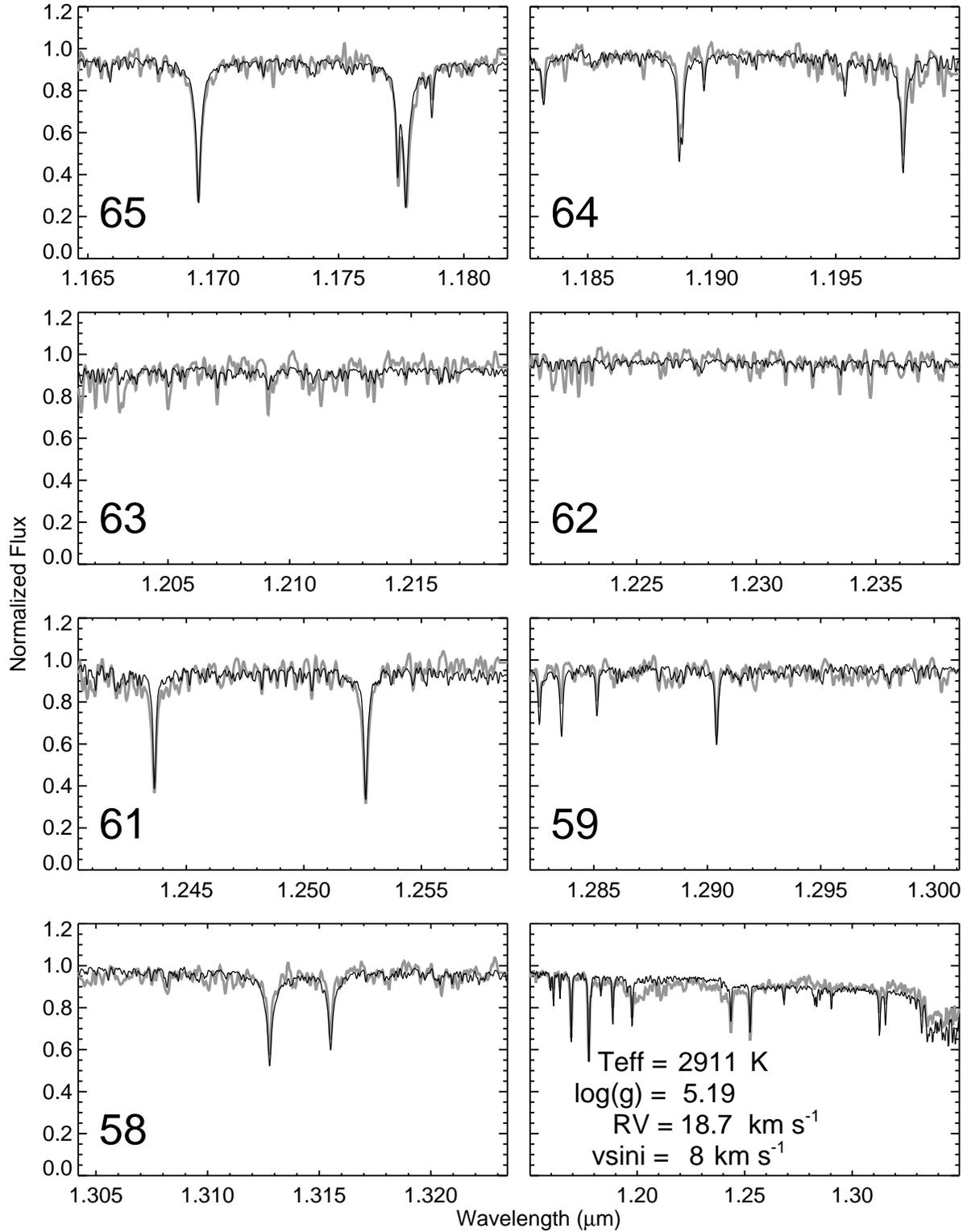}
   \caption{\label{gl-406fit} Observed (gray) and best-fit synthetic
     spectra (black) for the field (old, $>$1~Gyr) M6 dwarf
     Gl~406. The NIRSPEC dispersion orders are labeled, and the
     best-fit atmosphere model parameters are given in the bottom-right panel. Note the
     poor fit for the observed FeH features at both high- and
     medium-resolutions (see \S~\ref{sFeH}).
}
\end{figure}

\clearpage

\begin{figure}
  \includegraphics[height=.9\textheight,angle=0]{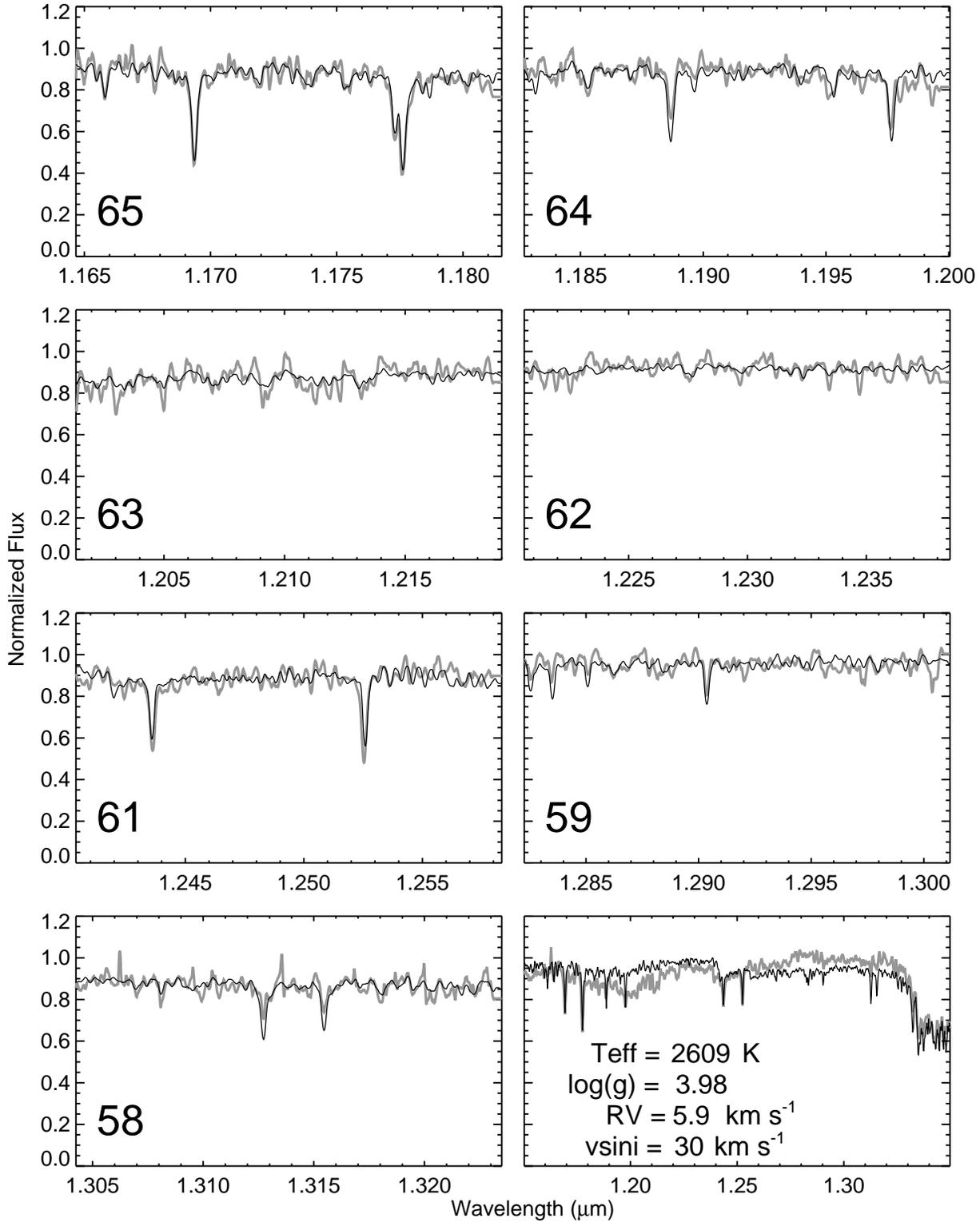}
   \caption{\label{2m1139fit} Observed (gray) and best-fit synthetic
     spectra for the young ($\sim$10~Myr) M8 TW Hydrae member
     2MASS~1139$-$31. The NIRSPEC dispersion orders are labeled, and
     the best-fit atmosphere model parameters are given in the bottom-right panel. Again,
     note the poor fit to FeH features (see \S~\ref{sFeH}).
}
\end{figure}

\clearpage

\begin{figure}
  \includegraphics[height=.9\textheight,angle=0]{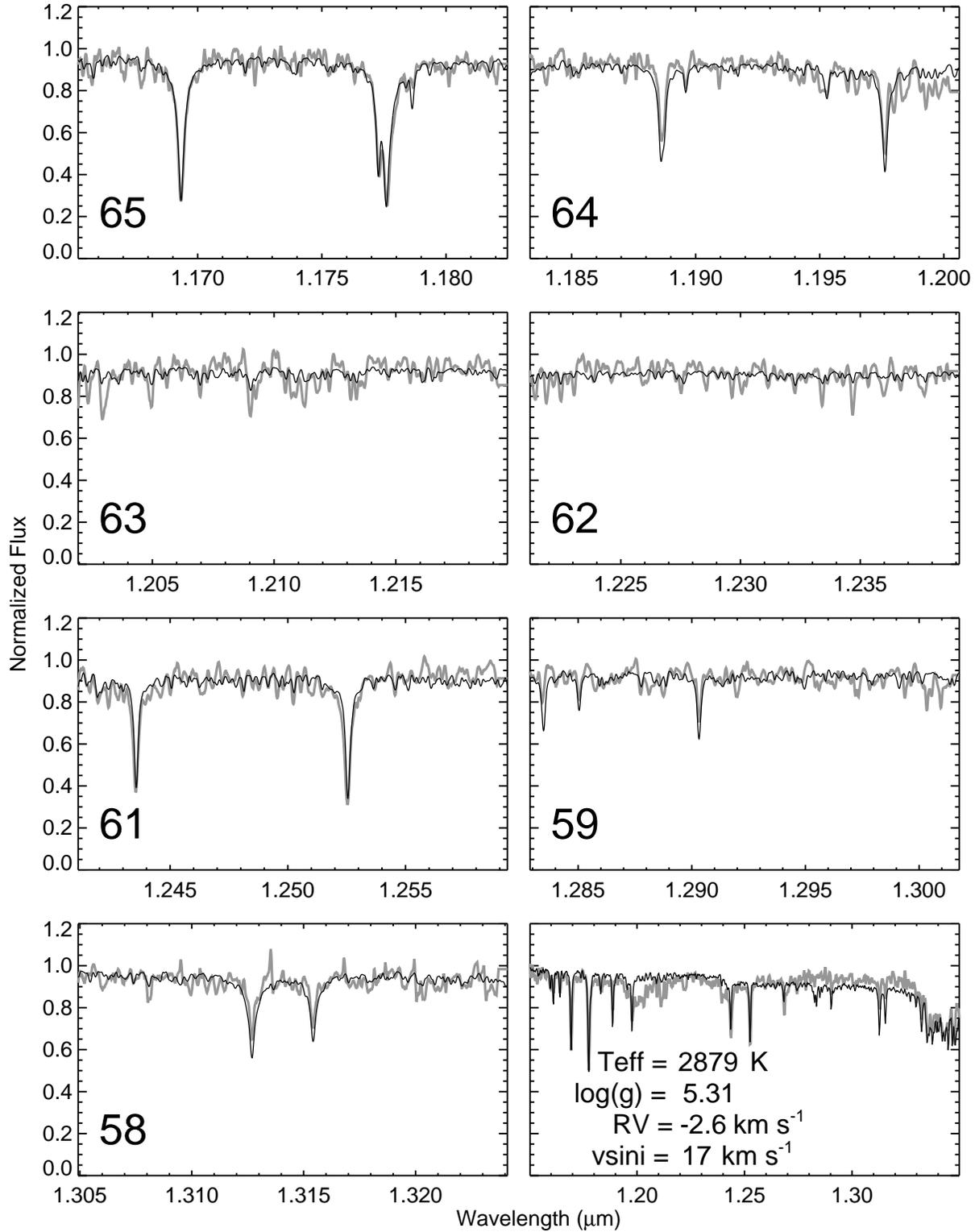}
   \caption{\label{l40258fit}Observed (gray) and best-fit synthetic
     spectra for the field (old, $>$1~Gyr) M7 dwarf LP~402-58. The
     NIRSPEC dispersion orders are labeled, and the best-fit atmosphere model
     parameters are given in the bottom-right panel.
}
\end{figure}

\clearpage

\begin{figure}
  \includegraphics[height=.8\textheight,angle=90]{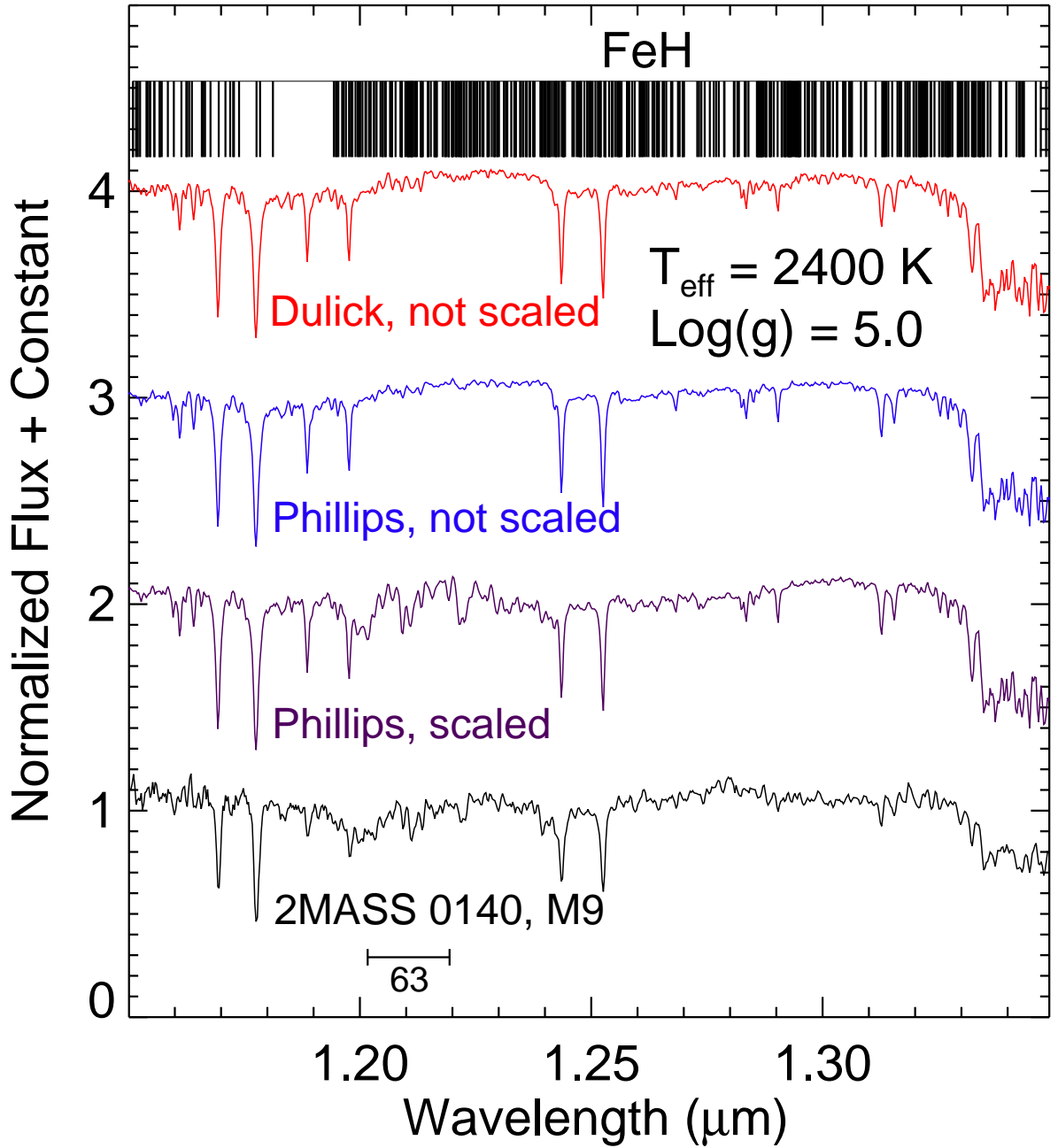}
   \caption{\label{fehmed} Synthetic spectra produced using the same
     atmospheric structure (T$_{eff}$=2400~K and log(g)=5.0) but
     different FeH line lists and oscillator-strength scalings as
     indicated on the plot. The NIRSPEC medium-resolution spectrum of
     the M9 dwarf 2MASS~0140+17 is plotted for comparison, and the
     horizontal bar marks the wavelength range of NIRSPEC order 63 shown in
     Figure~\ref{fehhi}. While the difference between the \citet{Dulick03} and
     the unscaled \citet{Phillips87} line lists are imperceptible at this
     resolution, the scaled \citet{Phillips87} oscillator strengths make a dramatic
     difference in the spectra from 1.205 to 1.240 $\mu$m, resulting in
     a much better match the the observed data. See \S~\ref{sFeH} for
     a detailed discussion.
}
\end{figure}

\clearpage

\begin{figure}
  \includegraphics[height=.8\textheight,angle=90]{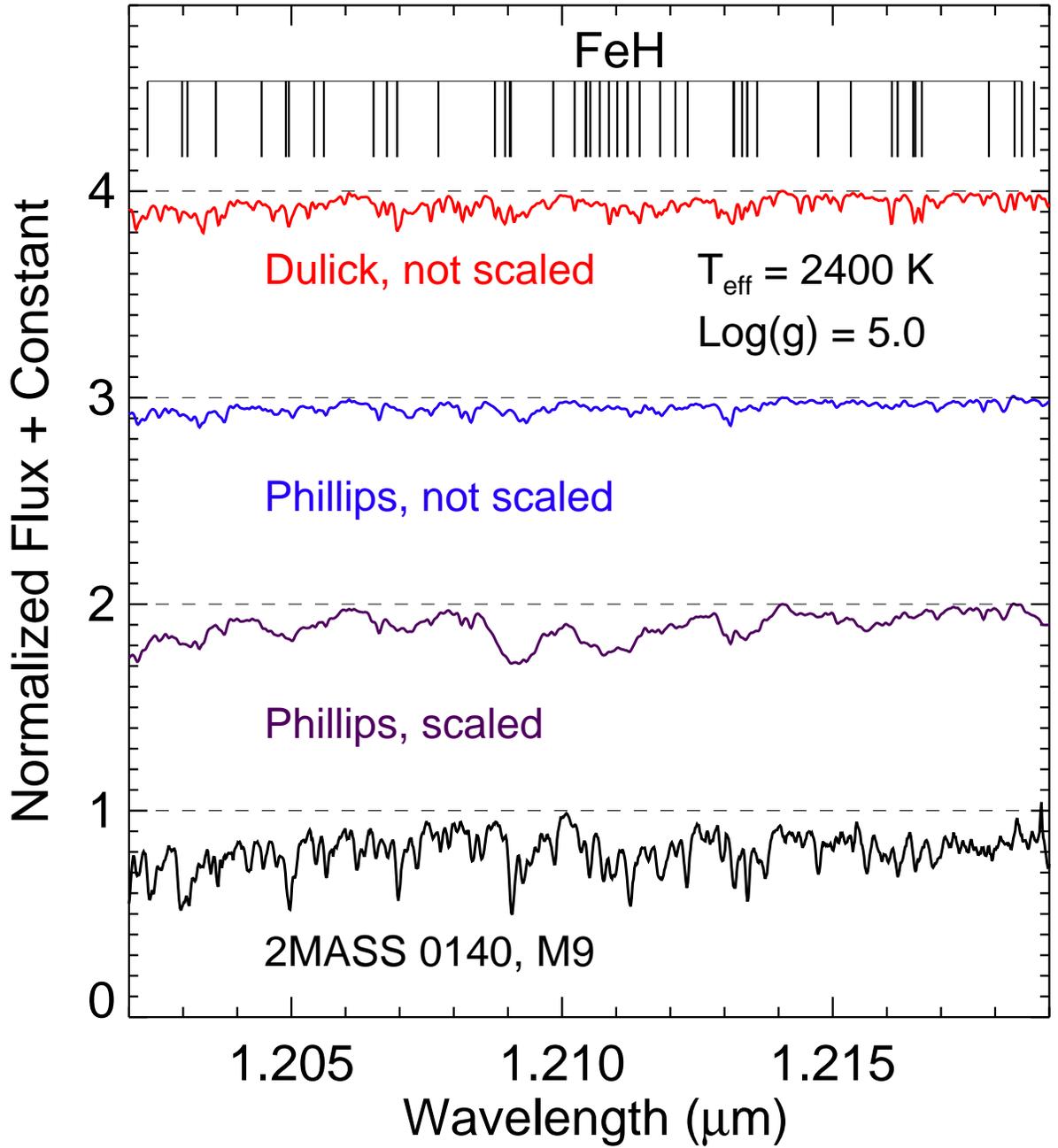}
   \caption{\label{fehhi} Synthetic spectra produced using the same
     atmospheric structure (T$_{eff}$=2400~K and log(g)=5.0) but
     different FeH line lists and oscillator-strength scalings as
     indicated on the plot. NIRSPEC dispersion order 63 spectrum of
     the M9 dwarf 2MASS~0140+27 is plotted for comparison. The
     synthetic spectrum using the \citet{Dulick03} line list shows an
     improved correspondence over the \citet{Phillips87} line list for
     individual features in the 2MASS~0140+27 spectrum, but the line
     strengths are too weak in the atmosphere model. See \S~\ref{sFeH} for a
     detailed discussion. 
}
\end{figure}

\clearpage

\begin{figure}
  \includegraphics[height=.8\textheight,angle=90]{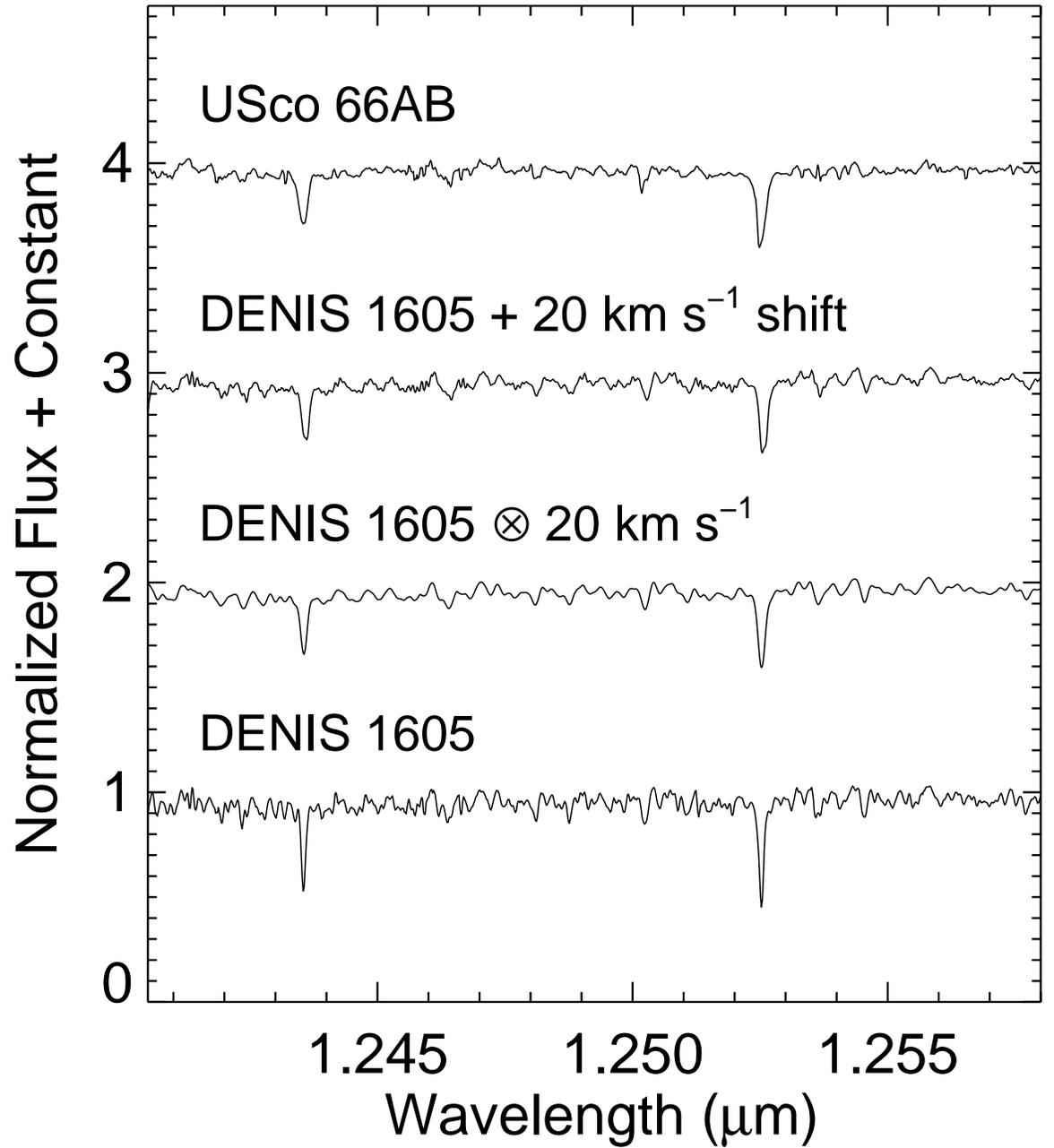}
   \caption{\label{binarytest} Observed NIRSPEC dispersion order~61
     spectra of  USco~66AB (top) and DENIS~1605$-$24 (bottom). The
     second-to-top spectrum are formed by shifting the spectrum of
     DENIS~1605$-$24 (log(g)=3.82, $v$sin$i$~$\le$~7~km~s$^{-1}$) by
     20 km~s$^{-1}$, the result added to the unshifted spectrum, and
     renormalizing. The second-from-bottom spectrum is the spectrum of
     DENIS~1605$-$24 convolved with a rotational velocity profile of
     20~km~s$^{-1}$. Both the shifted and convolved spectra resemble
     the spectrum of USco~66AB (log(g)=4.26, $v$sin$i$=28
     km~s$^{-1}$), indicating that binarity may influence the inferred
     $v$sin$i$ and/or log(g). This is unlikely to be the case for USco~66AB, for
     which the {\it known} binary would induce no more than a few
     km~s$^{-1}$ shift in the spectrum (assuming a circular orbit),
     but could be a factor in as yet unresolved objects such as
     DENIS~1619$-$24 and SCH~1622$-$19.
}
\end{figure}

\clearpage

\begin{figure}
  \includegraphics[height=.75\textheight,angle=90]{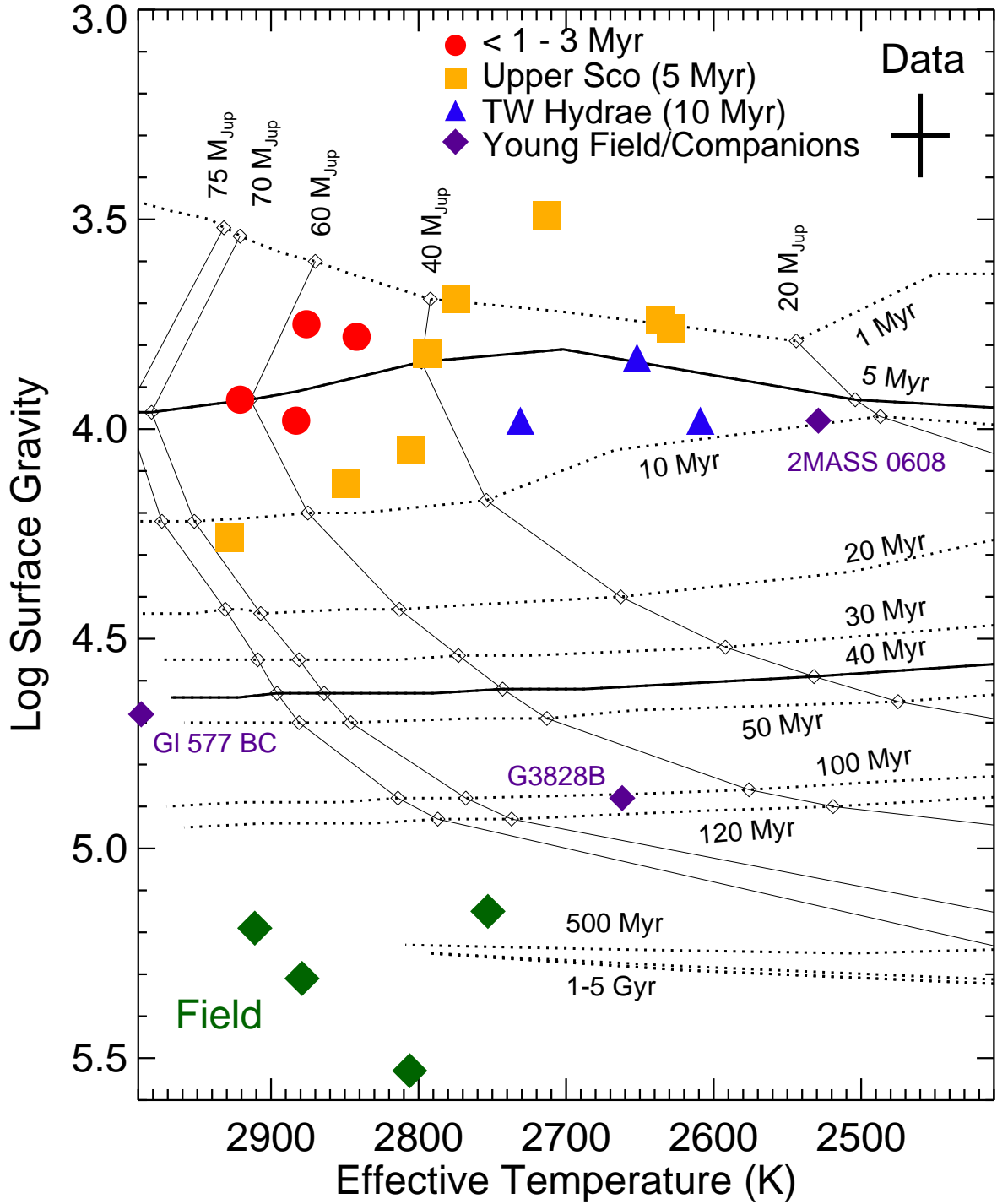}
   \caption{\label{evotracks2}Adopted physical properties of young
     objects (log(g)~$<$~4.3) plotted with the DUSTY00
     tracks from \citet{Chabrier00,Baraffe02}. Membership (and/or approximate age) are indicated by
     different symbols: circles for $\rho$ Ophiuchi, Taurus,
     LH$\alpha$233 group, and $\sigma$ Orionis objects; squares for Upper Scorpius objects;
     triangles for TW Hydrae objects, and diamonds for candidate young field objects and companions. 
     The error bars in the upper right represent
     typical uncertainties based on Monte Carlo simulations (see
     \S~\ref{fitting}); systematic uncertainties in the atmosphere models are
     likely larger.
}
\end{figure}

\clearpage

\begin{deluxetable}{llrcrrll}
\tabletypesize{\scriptsize}
\tablewidth{0pt} 
\tablecaption{\bf Observing Log 
\label{tbl-olog} }
\tablehead{ 
\colhead{Object} &
\colhead{Spec.} &
\colhead{R.A.} &
\colhead{Dec.} &
\colhead{$J$\tablenotemark{b}} &
\colhead{Int. Time} &
\colhead{UT Date} &
\colhead{$$} \\
\colhead{Name\tablenotemark{a}} &
\colhead{Type} &
\colhead{(J2000.0)} &
\colhead{(J2000.0)} &
\colhead{(mag)} &
\colhead{(seconds)} &
\colhead{of Observation\tablenotemark{c}} &
\colhead{Ref.}}
\startdata
Gl 577BC                & M5.5 & 15 05 49.9 &+64 02 50   & 7.19 & 1200 & 2006 May 19 &  1 \\
& & & & & 600 & 2003 Mar 24 & 3 \\
Gl 406 (Wolf 359)       & M6   & 10 56 28.9 &+07 00 53   &  7.09 & 240  & 2006 May 20 &  1 \\
& & & & & 120 & 2000 Dec 6 & 2 \\
CFHT Tau-7              & M5.75& 04 32 17.2 &+24 22 11   & 11.52 & 1200 & 2008 Mar 19 & 1 \\ 
& & & & & 600 & 2009 Nov 08 & 1 \\
$\sigma$ Orionis 12     & M6   & 05 37 57.5 &$-$02 38 44 & 14.23 & 6000 & 2008 Dec 06 & 1 \\
& & & & & 1200 & 2003 Dec 4 & 3 \\
UScoCTIO 66AB          & M6   & 16 01 49.6 &$-$23 51 08 & 12.91 & 2400 & 2007 May 29 & 1 \\
& & & & & 600 & 2003 May 12 & 3 \\
DENIS-P J160514.0-240652 & M6   & 16 05 14.0 &$-$24 06 52 & 12.84 & 2400 & 2007 May 29 & 1 \\
& & & & & 1200 & 2004 May 2 & 3 \\
GY 5 (ISO-Oph 30)       & M6   & 16 26 21.5 &$-$24 26 01 & 12.57 & 2400 & 2007 May 31 & 1 \\
& & & & & 600 & 2003 May 14 & 3 \\
2MASS J22344161+4041387AB    & M6   & 22 34 41.6 &+40 41 39	 & 12.57 & 2400 & 2007 May 29 & 1 \\
& & & & & 600 & 2003 Dec 4 & 3 \\
SCH J16121188-20472698   & M6.5 & 16 12 11.9 &$-$20 47 27 & 13.66 & 4800 & 2007 May 29 & 1 \\
& & & & & 1200 & 2009 Apr 7 & 1 \\
LP 402-58               & M7   & 23 36 43.9 &+21 53 39   & 12.71 & 3600 & 2008 Dec 7  & 1 \\
& & & & & 600 & 2004 Nov 6 & 3 \\
UScoCTIO 100            & M7   & 16 02 04.3 &$-$20 50 43 & 12.84 & 2400 & 2007 May 30 & 1 \\
& & & & & 600 & 2003 May 13 & 3 \\
UScoCTIO 130            & M7.5 & 15 59 43.7 &$-$20 14 40 & 14.20 & 4800 & 2007 May 30 & 1 \\
& & & & & 1200 & 2004 Apr 30 & 3 \\
LP 412-31               & M8   & 03 20 59.7 &+18 54 23   & 11.74 & 2400 & 2008 Dec 7  & 1 \\
& & & & & 600 & 2000 Dec 4 & 2 \\
SCH J16235158-23172740     & M8   & 16 23 51.7 &$-$23 17 26 & 13.55 & 3600 & 2008 Mar 20 & 1 \\ 
& & & & & 600 & 2009 Apr 7 & 1 \\
DENIS-P J161929.9-244047        & M8   & 16 19 29.9 &$-$24 40 47 & 14.23 & 2400 & 2007 May 30 & 1 \\
& & & & & 1200 & 2004 Jul 22 & 3 \\
SCH J16224384-19510575       & M8   & 16 22 43.8 &$-$19 51 06 & 12.35 & 2400 & 2008 Mar 19 & 1 \\
& & & & & 600 & 2009 Apr 7 & 1 \\
2MASS J11395113-3159214      & M8   & 11 39 51.1 &$-$31 59 21 & 12.69 & 3600 & 2006 May 19 & 1 \\
& & & & & 1200 & 2001 Dec 30 & 3 \\
2MASS J12073347-3932540AB   & M8   & 12 07 33.4 &$-$39 32 54 & 13.00 & 2400 & 2007 May 29 & 1 \\
& & & & & 1200 & 2001 Dec 30 & 3 \\
2MASS J06085283-2753583  & M8.5 & 06 08 52.8 &$-$27 53 58 & 13.60 & 4800 & 2008 Dec 6  & 1 \\ 
& & & & & 600 & 2003 Dec 4 & 3 \\
TWA 5B                  & M8.5   & 11 31 55.4 &$-$34 36 29 & 7.67 & 1200 & 2006 May 20 & 1 \\
& & & & & 600 & 2002 Dec 24 & 3 \\
2MASS J01400263+2701505      & M9   & 01 40 02.6 &+27 01 50   & 12.49 & 2400 & 2000 Dec 4  & 4  \\
& & & & & 600 & 2009 Nov 08 & 1 \\
\enddata
\tablenotetext{a}{2MASS, DENIS, and SCH object names are truncated in subsequent tables and in the text.}
\tablenotetext{b}{From 2MASS All-Sky Point Source Catalog. The magnitudes for the close companions Gl~577BC and TWA~5B are for the combined system because they are unresolved in 2MASS.} 
\tablenotetext{c}{The first entry for each object refers to high-resolution (echelle) observations and the second entry, if present, refers to medium-resolution (non-echelle) observations. References for these observations, if previously published, are given in the last column.}
\tablerefs{ (1) this work, (2) \citealt{McLean03}, (3) \citealt{McGovern05}, (4) \citealt{McLean07}} 
\end{deluxetable}

\begin{deluxetable}{lllc}
\tabletypesize{\scriptsize}
\tablewidth{0pt} 
\tablecaption{\bf Sample Properties
\label{samp} }
\tablehead{ 
\colhead{Object} &
\colhead{Spec.} &
\colhead{$$} &
\colhead{$$} \\
\colhead{Name} &
\colhead{Type} &
\colhead{Notes} &
\colhead{Ref. }}
\startdata
Upper Scorpius ($\sim$5~Myr) & & & \\
\hline
USco 66AB                 & M6   & q$\sim$1 binary               & 1, 2 \\
DENIS 1605$-$24           & M6   & \nodata                       & 3  \\
SCH 1612$-$20             & M6.5 & \nodata                       & 4 \\
USco 100                  & M7   & \nodata                       & 1  \\
USco 130                  & M7.5 & Estimated spectral type       & 1  \\
SCH 1623$-$23             & M8   & \nodata                       & 4  \\
DENIS 1619$-$24           & M8   & Possible spectroscopic binary & 3, 4  \\
SCH 1622$-$19             & M8   & Possible unresolved binary    & 5  \\
\hline
TW Hydrae ($\sim$10~Myr) & & & \\
\hline
2MASS 1139$-$31           & M8   & \nodata                      & 6 \\
2MASS 1207$-$39AB         & M8   & q$\sim$0.3 binary            & 6, 7 \\
TWA 5B                    & M8.5 & Companion                    & 8 \\
\hline
Other Young Objects & & & \\
\hline
GY 5 (ISO-Oph 30)         & M6   &$\rho$~Ophiuchi, $\lesssim$1~Myr                & 9 \\
2MASS 2234+40AB           & M6   &Lk~H$\alpha$~233,  $\sim$1~Myr, q$\sim$1 binary & 10 \\
CFHT Tau-7                & M5.75, M6.5 & Taurus, 1.5~Myr                         & 11, 12 \\
$\sigma$ Ori 12           & M6   &$\sigma$~Orionis, $\sim$3~Myr                   & 13 \\
Gl 577BC                  & M5.5 & Companion, $\sim$70~Myr, q$\sim$1 binary       & 14 \\
2MASS 0608$-$27           & M8.5 & Young Field, $\le$100~Myr                      & 15  \\
\hline
Field Objects ($\ge$1~Gyr) & & & \\
\hline
Gl 406 (Wolf 359)         & M6   & RV standard                  & 16  \\
LP 402-58                 & M7   & \nodata                      & 17 \\
LP 412-31                 & M8   & Strongly flaring object      & 17, 18 \\
2MASS 0140+27             & M9   & \nodata                      & 16 \\
\enddata
\tablerefs{(1) \citet{Ardila00}, (2) \citet{Kraus05}, (3) \citet{Martin04}, (4) \citep{Mohanty05}(5) \citet{Slesnick06}, (6) \citet{Gizis02}, (7) \citet{Mohanty07}, (8) \citet{Lowrance99}, (9) \citet{Wilking99}, (10) \citet{Allers09}, (11)\citet{Luhman06a}, (12) \citet{Guieu06}, (13) \citet{Bejar99}, (14) \citet{Lowrance05}, (15) \citet{Kirkpatrick08}, (16) \citet{McLean07}, (17) \citet{Gizis00}, (18) \citet{Schmidt07}. } 
\end{deluxetable}

\begin{deluxetable}{lclcccccccccc}
\tabletypesize{\scriptsize}
\tablewidth{0pt} \tablecaption{\bf Spectral Fitting Results - Effective Temperatures
\label{tbl-10myr-temp}} 
\tablehead{ 
\colhead{$ $} &
\colhead{Spec.} &
\colhead{Prev. } &
\colhead{} &
\colhead{} &
\colhead{} &
\colhead{} &
\colhead{} &
\colhead{} &
\colhead{} &
\colhead{} &
\colhead{Adopted} \\
\colhead{Object} &
\colhead{Type} &
\colhead{T$_{eff}$\tablenotemark{a}} &
\colhead{N3 } &
\colhead{65 } &
\colhead{64 } &
\colhead{63 } &
\colhead{62 } &
\colhead{61 } &
\colhead{59 } &
\colhead{58 } &
\colhead{T$_{eff}$ }}
\startdata
USco 66AB               & M6  & 2851--3000  &  2973   &  2927  &   2939    &  (2800)  &\sout{2250}&   2988   &  (3000)  &  (2881)  &  2928 \\
DENIS 1605$-$24         & M6  & 2850--3000  &  2826   &  2785  &   2739    &  (2537)  &  (2800)  &  (2977)  &   2921   &   2800   &  2794 \\
SCH 1612$-$20           & M6.5& 2630--2935  &  2840   &  2804  &   2757    &   2531   &   2699   &  (3000)  &   2999   &   2624   &  2777 \\
USco 100                & M7  & 2672--2880  &  2880   &  2825  &   2813    & -(2651)- & -(2998)- &  (2912)  &  (3000)  &  (2746)  &  2849 \\
USco 130                & M7.5& 2550--2850  &  2660   &  2653  &   2764    &  (2800)  &  (2883)  &  (2826)  &  (2978)  &  (2741)  &  2805 \\
SCH 1623$-$23           & M8  & 2455--2710  &  2625   & (2489) &  (2780)   &  (2351)  & -(2458)- &  (2741)  & -(2899)- & -(2664)- &  2636 \\
DENIS 1619$-$24         & M8  & 2600--2710  &  2726   & (2666) &  (2770)   &\sout{2645}&\sout{2906}&  (2776)  &  (2810)  &  (2558)  &  2713 \\
SCH 1622$-$19           & M8  & 2400--2710  &  2594   &  2594  &  (2632)   &  (2261)  &  (2376)  &  (2800)  & -(2939)- & -(2792)- &  2629 \\
2MASS 1139$-$31         & M8  & 2400--2710  & (2435)  &  2486  &  (2701)   &  (2409)  & -(2900)- &  (2600)  & -(2837)- &  (2484)  &  2609 \\
2MASS 1207$-$39AB       & M8  & 2400--2710  &  2485   &  2602  &   2694    &   2405   &  (2800)  &  (2801)  &  (2804)  &   2486   &  2652 \\
2MASS 0608$-$27         & M8.5& 2150--2555  &  2200   & (2574) & -(2585)-  &  (2167)  &   2276   & -(2900)- & -(2727)- & -(2474)- &  2529 \\
TWA 5B                  & M8.5& 2300--2550  &  2360   & (2716) &  (2686)   &  (2391)  &   2787   &\sout{2620}&  (2904)  &  (2814)  &  2731 \\
\hline
Gl 406                  & M6   & 2670--3058 &  3000  &  (3000)  &  (2833)  &\sout{2315}&  (2895)  &  (2900)  &\sout{3000}&\sout{3000}&  2911 \\
LP 402-58               & M7   & 2600--2620 &  2864  &  (2950)  & -(2800)- &  (2900)  &  (2799)  &  (2900)  &\sout{3000}&\sout{2960}&  2879 \\
LP 412-31               & M8   & 2480--2638 &  2919  &  (2936)  &  (2950)  & -(2273)- & -(2805)- &  (2860)  &\sout{2800}&\sout{2600}&  2806 \\
2MASS 0140+27           & M9   & 2325--2400 &  2824  &  (2805)  &  (2850)  & -(2215)- & -(2846)- & -(2900)- &\sout{3000}&\sout{2714}&  2753 \\
\enddata
\tablenotetext{a}{Effective temperature (or range of temperatures) determined by previous studies for the specific object or for an object of the same spectral type and similar age are from \citet{Basri00,Dahn02,Luhman03,Gorlova03,Mohanty03b,Golimowski04,Mohanty04a,Vrba04,McGovern05,Reiners07b,Slesnick06,ScholzA07,Slesnick08,Teixeira08,Wilking99}. }
\tablecomments{See \S~\ref{fitting} for explanation of annotations of the table entries.}
\end{deluxetable}

\begin{deluxetable}{lclccccccccc}
\tabletypesize{\scriptsize}
\tablewidth{0pt} \tablecaption{\bf Spectral Fitting Results - Surface Gravities
\label{tbl-10myr-logg}} 
\tablehead{ 
\colhead{$ $} &
\colhead{ }&
\colhead{Prev.} &
\colhead{} &
\colhead{} &
\colhead{} &
\colhead{} &
\colhead{} &
\colhead{} &
\colhead{} &
\colhead{} &
\colhead{Adopted} \\
\colhead{Object} &
\colhead{Age\tablenotemark{a} } &
\colhead{Log(g)\tablenotemark{b} } &
\colhead{N3 } &
\colhead{65 } &
\colhead{64 } &
\colhead{63 } &
\colhead{62 } &
\colhead{61 } &
\colhead{59 } &
\colhead{58 } &
\colhead{log(g) }}
\startdata
USco 66AB               & 5~Myr      & 3.64--4.50 &  4.21   &   4.32   &   4.20   &  (4.69)  &\sout{5.40}&   4.72   &  (3.80)  &  (3.92)   &  4.26 \\
DENIS 1605$-$24         & 5~Myr      & 4.25       &  3.70   &   3.60   &   3.49   &  (3.60)  &  (4.47)  &  (4.67)  &   3.94   &   3.14    &  3.82 \\
SCH 1612$-$20           & 5~Myr      & 3.57--3.61 &  3.50  &   3.66   &   3.46   &   3.81   &   3.53   &  (4.67)  &   3.80   &   3.01    &  3.68 \\
USco 100                & 5~Myr      & 3.55--4.50 &  4.10   &   4.09   &   3.82   & -(4.70)- & -(5.69)- &  (4.75)  &  (3.72)  &  (3.09)   &  4.13 \\
USco 130                & 5~Myr      & 3.25--4.00 &  3.82   &   3.92   &   3.56   &  (4.70)  &  (4.35)  &  (4.67)  &  (3.99)  &  (3.12)   &  4.05 \\
SCH 1623$-$23           & 5~Myr      & 3.27--3.32 &  3.66   &  (3.57)  &  (3.48)  &  (3.89)  & -(4.41)- &  (4.46)  & -(3.61)- & -(3.00)-  &  3.74 \\
DENIS 1619$-$24         & 5~Myr      & 4.00       &  3.52   &  (3.66)  &  (3.38)  &\sout{5.27}&\sout{5.76}&  (4.19)  &  (3.21)  &  (3.00)   &  3.49 \\
SCH 1622$-$19           & 5~Myr      & 3.02--3.06 &  3.70   &   3.89   &  (3.38)  &  (3.79)  &  (3.56)  &  (4.77)  & -(3.85)- & -(3.01)-  &  3.76 \\
2MASS 1139$-$31         & 10~Myr     & 3.62--4.25 & (3.30)  &   3.76   &  (3.82)  &  (4.03)  & -(5.59)- &  (4.30)  & -(3.57)- &  (3.33)   &  3.98 \\
2MASS 1207$-$39AB       & 10~Myr     & 4.00--4.25 &  3.30   &   3.90   &   3.50   &   3.66   &  (4.60)  &  (4.65)  &  (3.40)  &   3.00    &  3.83 \\
2MASS 0608$-$27         & $<$100~Myr & 4.00       &  3.20   &  (4.42)  & -(3.61)- &  (3.92)  &   3.60   & -(5.57)- & -(3.40)- & -(3.00)-  &  3.98 \\
TWA 5B                  & 10~Myr     & 3.75--3.98 &  3.10   &  (4.12)  &  (3.50)  &  (4.20)  &   4.34   &\sout{5.26}&  (3.60)  &  (3.53)   &  3.88 \\
\hline
Gl 406                  & $>$1 Gyr   & 5.15--5.40 &  5.67   &  (5.61)  &  (4.56)  &\sout{3.62}&  (4.85)  &  (5.60)  &\sout{4.15}&\sout{4.31}&  5.19  \\
LP 402-58               & $>$1 Gyr   & 5.15--5.40 &  5.35   &  (5.55)  & -(4.70)- &  (5.60)  &  (4.56)  &  (5.73)  &\sout{4.00}&\sout{3.99} & 5.31 \\
LP 412-31               & $>$1 Gyr   & 5.15--5.40 &  6.00   &  (6.00)  &  (6.00)  & -(3.88)- & -(4.79)- &  (6.00)  &\sout{3.60}&\sout{4.20} & 5.53  \\
2MASS 0140+27           & $>$1 Gyr   & 5.15--5.40 &  5.83   &  (5.51)  &  (5.10)  & -(3.61)- & -(4.88)- & -(6.00)- &\sout{3.71}&\sout{3.40} & 5.15  \\
\enddata
\tablenotetext{a}{Age references: \citealt{Gizis00,Preibisch02,Zuckerman04,BarNav06,Kirkpatrick08} }
\tablenotetext{b}{Surface gravities determined by previous studies for young objects are from \citet{Gorlova03,Mohanty04a,McGovern05,BarNav06,Slesnick06,Slesnick08,Teixeira08}. The range of surface gravity for field objects (age~$>$1~Gyr) is set by evolutionary models of \citet{Chabrier00} and \citet{Baraffe02}. }
\tablecomments{See \S~\ref{fitting} for explanation of annotations of the table entries.}
\end{deluxetable}

\begin{deluxetable}{lclccccccccc}
\tabletypesize{\footnotesize}
\tablewidth{0pt} \tablecaption{\bf M6 Sequence - Effective Temperatures
\label{tbl-M6age-temp}} 
\tablehead{ 
\colhead{$ $} &
\colhead{Spec.} &
\colhead{Prev.\tablenotemark{a} } &
\colhead{} &
\colhead{} &
\colhead{} &
\colhead{} &
\colhead{} &
\colhead{} &
\colhead{} &
\colhead{} &
\colhead{Adopted} \\
\colhead{Object} &
\colhead{Type} &
\colhead{T$_{eff}$\tablenotemark{a}} &
\colhead{N3 } &
\colhead{65 } &
\colhead{64 } &
\colhead{63 } &
\colhead{62 } &
\colhead{61 } &
\colhead{59 } &
\colhead{58 } &
\colhead{T$_{eff}$ }}
\startdata
GY 5                    & M6\tablenotemark{b} & 2700--3050   &  2871  &  2939  &  2814  &  (2676)  &  (2788)  &  (3000)  &  (3000)  &   2845   & 2876 \\
2MASS 2234+40AB         & M6                  & 2990         &  2938  &  2950  &  2929  &  (3000)  &  (3000)  &   3000   &   2913   &   2617   & 2921   \\
CFHT Tau-7              & M6\tablenotemark{c} & 2935--3024   &  2740  &  2805  &  2801  &  (2701)  &  (2775)  &   2946   &   2969   &   2760   & 2825 \\
$\sigma$Ori 12          & M6                  & 2990         &  2779  &  2973  &  2833  &   2650   &   2700   &   3000   &   2939   &   2793   & 2842 \\
DENIS 1605$-$24         & M6                  & 2850--3000   &  2826  &  2985  &  2739  &  (2537)  &  (2800)  &  (2977)  &   2921   &   2800   & 2794 \\
Gl 577BC                & M5.5                & 2900--292    &  3000  &  3000  & (2961) & -(3000)- &  (2970)  &  (2999)  &  (3000)  &  (3000)  & 2988 \\
Gl 406                  & M6                  & 2670--3058   &  3000  & (3000) & (2833) &\sout{2315}&  (2895) &  (2900)  &\sout{3000}&\sout{3000}& 2911 \\
\enddata
\tablenotetext{a}{Effective temperature (or range of temperatures) determined by previous studies for the specific object or for an object of the same spectral type and similar age are from \citet{Luhman99,Luhman03,Lowrance05,Mohanty05,Wilking05,Gatti06,Guieu06}. }
\tablenotetext{b}{Classified as M5.5 by \citet{Wilking05} }
\tablenotetext{c}{Classified as M5.75 by \citet{Luhman06a} and as M6.5 by \citet{Guieu06} }
\tablecomments{See \S~\ref{fitting} for explanation of annotations of the table entries.}
\end{deluxetable}

\begin{deluxetable}{lclccccccccc}
\tabletypesize{\footnotesize}
\tablewidth{0pt} \tablecaption{\bf M6 Sequence - Surface Gravities
\label{tbl-M6age-logg}} 
\tablehead{ 
\colhead{$ $} &
\colhead{Age} &
\colhead{Prev. } &
\colhead{} &
\colhead{} &
\colhead{} &
\colhead{} &
\colhead{} &
\colhead{} &
\colhead{} &
\colhead{} &
\colhead{Adopted} \\
\colhead{Object} &
\colhead{ } &
\colhead{Log(g)} &
\colhead{N3 } &
\colhead{65 } &
\colhead{64 } &
\colhead{63 } &
\colhead{62 } &
\colhead{61 } &
\colhead{59 } &
\colhead{58 } &
\colhead{Log(g) }}
\startdata
GY 5                    &$\lesssim$1 Myr & 3.65       & 3.80  &  3.57  &  3.30  &  (4.14)  &  (4.08)  & (4.39) &  (3.80)  &   3.24   & 3.75 \\
2MASS 2234+40AB         & $\sim$1 Myr    & 4.00       & 3.80  &  3.54  &  3.56  &  (5.00)  &  (4.81)  &  4.19  &   3.69   &   3.60   & 3.93 \\
CFHT Tau-7              &   1.5 Myr      & \nodata    & 3.50  &  3.40  &  3.44  &  (4.00)  &  (4.16)  &  4.32  &   3.80   &   3.10   & 3.72 \\
$\sigma$Ori 12          &     3 Myr      & 4.00       & 3.70  &  4.05  &  3.59  &   3.61   &   3.74   &  4.49  &   3.78   &   3.21   & 3.78 \\
DENIS 1605$-$24         & 5~Myr          & 4.25       & 3.70  &  3.60  &  3.49  &  (3.60)  &   4.47   &  4.67  &   3.94   &   3.14   & 3.82 \\
Gl 577BC                &  70~Myr        & 4.75       & 4.43  &  4.87  & (4.56) & -(5.59)- &  (4.87)  & (5.09) &  (3.80)  &  (4.07)  & 4.68 \\
Gl 406                  & $>$1 Gyr       & 5.15--5.40 & 5.67  & (5.61) & (4.56) &\sout{3.62}&  (4.85)  &  5.60  &\sout{4.15}&\sout{4.31}& 5.19  \\
\enddata
\tablerefs{Age and gravity from \citet{Luhman99,Lowrance05,McGovern05,Mohanty05,Wilking05,Caballero07,Allers09}. }
\tablecomments{See \S~\ref{fitting} for explanation of annotations of the table entries.}
\end{deluxetable}

\begin{deluxetable}{lcrrrr}
\tabletypesize{\footnotesize}
\tablewidth{0pt} \tablecaption{\bf Velocity Results
\label{tbl-vel} }
\tablehead{ 
\colhead{} &
\colhead{} &
\colhead{~~~~Previous} &
\colhead{Results~~~~} &
\colhead{~~~~~Current} &
\colhead{Results~~~~}  \\
\colhead{$ $} &
\colhead{Spectral} &
\colhead{RV} &
\colhead{$v$~sin$i$} &
\colhead{RV} &
\colhead{$v$~sin$i$}  \\
\colhead{Object} &
\colhead{Type} &
\colhead{(km~s$^{-1}$) } &
\colhead{(km~s$^{-1}$) } &
\colhead{(km~s$^{-1}$) } &
\colhead{(km~s$^{-1}$) } }
\startdata
Upper Scorpius & & & & & \\
\hline
USco 66AB               & M6   &$-$7.81,$-$4.4&25.9, 27.5 & $-$5.8~~ & 28~~~~ \\
DENIS 1605$-$24         & M6   & \nodata      &  \nodata  & $-$5.2~~ &$\le$7~~~~ \\
SCH 1612$-$20           & M6.5 & \nodata      &  \nodata  & $-$6.8~~ &$\le$9~~~~ \\
USco 100                & M7   & $-$7.2,$-$8.9& 43.7, 50  & $-$8.3~~ & 50~~~~ \\
USco 130                & M7.5 &  $-$3.91     & 14, 15.2  & $-$7.1~~ & 18~~~~ \\
SCH 1623$-$23           & M8   & \nodata      &  \nodata  & $-$6.0~~ & 53~~~~ \\
DENIS 1619$-$24         & M8   & \nodata      &  \nodata  & $-$7.0~~ & 47~~~~ \\
SCH 1622$-$19           & M8   & \nodata      &  \nodata  &$-$10.2~~ & 25~~~~ \\
\hline
TW Hydrae & & & & & \\
\hline
2MASS 1139$-$31         & M8   &  11.6, 9.7   &   25      &    5.9~~ & 30~~~~ \\
2MASS 1207$-$39AB       & M8   &  11.2, 8.7   &   13      &    7.5~~ & 18~~~~ \\
TWA 5B                  & M8.5 &  13.4        &   16      &    9.7~~ & 27~~~~ \\
\hline
Other Young Objects & & & & & \\
\hline
2MASS 2234+40AB          & M6   &  $-$10.6     &  \nodata  &$-$13.1~~ & 17~~~~ \\
CFHT Tau-7              & M6.5 & \nodata      &  \nodata  &   15.6~~ & 20~~~~ \\
$\sigma$ Ori 12         & M6   & 29.8, 37     &  \nodata  &   31.2~~ & 18~~~~ \\
GY 5                    & M6   &$-$6.39,$-$6.3&16.5, 16.8 & $-$7.1~~ & 20~~~~ \\
Gl 577BC                & M5.5 &  \nodata     &  \nodata  & $-$6.2~~ & 43~~~~ \\
2MASS 0608$-$27         & M8.5 &  \nodata     & \nodata   &   22.6~~ & 20~~~~ \\
\hline
Field Objects & & & & & \\
\hline
Gl 406                  & M6   &   19.1       &  $\le$2.9 &   18.7~~ &$\le$8~~~~ \\
LP 402-58               & M7   &  \nodata     &   \nodata & $-$2.6~~ & 17~~~~ \\
LP 412-31               & M8   & 41.6, 44.7   &   8,9     &   41.8~~ & 33~~~~ \\
2MASS 0140+27           & M9   & 9.6, 8.2     &  6.5      &    8.6~~ & 11~~~~ \\
\enddata
\tablerefs{\citet{Tinney98,Reid02,Reid03,Mohanty03,Mohanty03b,Muzerolle03,Kenyon05,Mohanty05,Kurosawa06,Reiners07b,Allers09}.}
\end{deluxetable}

\end{document}